\documentclass[twocolumns]{aa}
\usepackage{graphicx}
\usepackage{natbib}
\bibpunct{(}{)}{;}{a}{}{,}
\usepackage{wasysym}
\let\iint\undefined 
 
\usepackage{amsmath}
\usepackage{cancel}
\usepackage{multirow}
\usepackage{float}

\begin{document}

\title{Deep probing of the photospheric sunspot penumbra: no evidence for magnetic field-free gaps}
\author{J.M.~Borrero\inst{1} \and A.~Asensio Ramos\inst{2,3} \and M.~Collados\inst{2,3} \and
R.~Schlichenmaier\inst{1} \and H.~Balthasar\inst{4} \and M.~Franz\inst{1} \and R.~Rezaei\inst{1,2,3}
\and C.~Kiess\inst{1} \and D.~Orozco Su\'arez\inst{2,3} \and A.~Pastor\inst{2,3} \and T.~Berkefeld\inst{1} 
\and O.~von der L\"uhe\inst{1} \and D.~Schmidt\inst{1} \and W.~Schmidt\inst{1} \and M.~Sigwarth\inst{1} 
\and D.~Soltau\inst{1} \and R.~Volkmer\inst{1} \and T.~Waldmann\inst{1} \and C.~Denker\inst{4} \and 
A.~Hofmann\inst{4} \and J.~Staude\inst{4} \and K.G.~Strassmeier\inst{4} \and A.~Feller\inst{5} \and A.~Lagg\inst{5} \and S.K.~Solanki\inst{5,8} \and M.~Sobotka\inst{7} \and H.~Nicklas\inst{6}}
\institute{Kiepenheuer-Institut f\"ur Sonnenphysik, Sch\"oneckstr. 6, D-79110, Freiburg, Germany
\and
Instituto de Astrof{\'\i}sica de Canarias, Avd. V{\'\i}a L\'actea s/n, E-38205, La Laguna, Spain
\and
Departamento de Astrof{\'\i}sica, Universidad de La Laguna, E-38205, La Laguna, Tenerife, Spain
\and
Leibniz Institute for Astrophysics Potsdam, An der Sternwarte 16, D-14482, Potsdam, Germany
\and
Max Planck Institut for Solar System Reesearch, Justus-von-Liebig-Weg 3, D-37077, G\"ottingen, Germany
\and
Georg-August-Universit\"at G\"ottingen, Faculty of Physics, Friedrich-Hund-Platz 1, D-37077, G\"ottingen, Germany
\and
Astronomical Institute, Academy of Sciences of the Czech Republic, Fri\v{c}ova 298, 25165 Ond\v{r}ejov, Czech Republic
\and
School of Space Research, Kyung Hee University, Yongin, 446-701, Gyeonggi, Republic of Korea
}
\date{Recieved / Accepted}

\abstract{Some models for the topology of the magnetic field in sunspot penumbrae predict the existence of 
field-free or dynamically weak-field regions in the deep Photosphere.}{To confirm or rule out the existence 
of weak-field regions in the deepest photospheric layers of the penumbra.}{The magnetic
field at $\log\tau_5=0$ is investigated by means of inversions of spectropolarimetric data of two different sunspots
located very close to disk center with a spatial resolution of approximately 0.4-0.45\arcsec. The data have been recorded 
using the GRIS instrument attached to the 1.5-meters GREGOR solar telescope at El Teide observatory. It includes 
three Fe \textsc{i} lines around 1565 nm, whose sensitivity to the magnetic field peaks at half a pressure-scale-height deeper
than the sensitivity of the widely used Fe \textsc{i} spectral line pair at 630 nm. Prior to the inversion, the data is 
corrected for the effects of scattered light using a deconvolution method with several point spread functions.}{At 
$\log\tau_5=0$ we find no evidence for the existence of regions with dynamically weak ($B<500$~Gauss) magnetic fields 
in sunspot penumbrae. This result is much more reliable than previous investigations done with Fe \textsc{i} lines at 630 nm.
Moreover, the result is independent of the number of nodes employed in the inversion, and also independent of the point 
spread function used to deconvolve the data, and does not depend on the amount of straylight (i.e. wide-angle scattered light) 
considered.}{}

\titlerunning{Deep-probing of sunspot penumbra: no evidence for field-free gaps}
\authorrunning{Borrero et al.}
\maketitle

\def\kms{~km s$^{-1}$}
\def\deg{^{\circ}}
\def\df{{\rm d}}
\newcommand{\ve}[1]{{\rm\bf {#1}}}
\newcommand{\diff}{{\rm d}}
\newcommand{\Conv}{\mathop{\scalebox{1.5}{\raisebox{-0.2ex}{$\ast$}}}}%
\def\ex{{\bf e_x}}
\def\ez{{\bf e_z}}
\def\ey{{\bf e_y}}
\def\expr{{\bf e_x^\ensuremath{\prime}}}
\def\ezpr{{\bf e_z^\ensuremath{\prime}}}
\def\eypr{{\bf e_y^\ensuremath{\prime}}}
\def\xp{x^\ensuremath{\prime}}
\def\yp{y^\ensuremath{\prime}}
\def\xas{x^{\ast}\!}
\def\yas{y^{\ast}\!}
\def\C{\mathcal{C}}

\section{Introduction}%
\label{section:intro}

The last decade has been witness to an unprecedented advance in our knowledge
of sunspot penumbrae. Owing to the improvement in instrumentation, data
analysis methods and realism of numerical simulations, an unified picture for the topology of 
the penumbral magnetic and velocity fields has begun to emerge. The foundations
of this picture rest on the so-called \emph{spine/intraspine} structure of the
sunspot penumbra, first mentioned by \citet{lites1993pen}, whereby regions of strong
and somewhat vertical magnetic fields (i.e. spines) alternate horizontally with regions of weaker
and more inclined field lines that harbor the Evershed flow (i.e. intraspines). At low spatial
resolution ($\approx 1\arcsec$) the intraspines are identified with penumbral filaments. At the same time,
\citet{solanki1993pen} established that these two distinct components also interlace vertically,
thereby explaining the asymmetries in the observed circular polarization profiles (Stokes $V$). It was later
found that the vertical and horizontal interlacing of these two components implies that the magnetic 
field in the spines wraps around the intraspines \citep{borrero2008pen}, with the latter remaining 
unchanged at all radial distances from the sunspot's center \citep{borrero2005pen,borrero2006pen,tiwari2013decon} 
and the former being nothing but the extension of the umbral field into the penumbra \citep{tiwari2015decon}.
It has also been confirmed that the Evershed flow can reach supersonic and super-Alv\'enic values, not only
on the outer penumbra \citep{borrero2005pen,vannoort2013decon}, but also close to the umbra \citep{deltoro2001pen,bellot2004pen}
and has a strong upflowing component at the inner penumbra that turns into a downflowing component at larger
radial distances \citep{franz2009pen,franz2013pen,tiwari2013decon}. Finally, there is strong evidence for an additional 
component of the velocity field in intraspines that appears as convective upflows along the center of the intraspines
that turns into downflows at the filaments' edges \citep{zakharov2008pen,joshi2011pen,
scharmer2011pen,tiwari2013decon}. These downflows seem capable of dragging the magnetic field lines and turning 
them back into the solar surface \citep{basilio2013pen,scharmer2013pen}.

In spite of this emerging unified picture, a number of controversies persist. One of them pertains to the
strength of the convective upflows/downflows at the intraspines' center/edges. \citet{tiwari2013decon,pozuelo2015pen}
found an average speed for this convective velocity pattern of about 200 m\,s$^{-1}$. Although they are ubiquitous, their
strength does not seem capable of sustaining the radiative cooling of the penumbra, which amounts to about 
70 \% of the quiet-Sun brightness. However, \citet{scharmer2013pen} find an rms convective velocity of 1.2 km\,s$^{-1}$ 
at the intraspines' center/edges, hence strong enough to explain the penumbral brightness. The latter result
agrees well with numerical simulations of sunspot penumbrae \citep{rempel2012mhd}. On the other hand, the pattern 
of upflows/downflows at the heads/tails, respectively, at the penumbral intraspines is easily discernible
\citep{franz2009pen,ichimoto2010pen, franz2013pen} and harbors plasma flows of several km\,s$^{-1}$, albeit occupying only a 
small fraction of the penumbral area. Which one of these two aforementioned convective modes accounts for the energy 
transfer in the penumbra is unclear from an observational point of view, although the scale is starting to tip in favor
of the former.

Another remaining controversy concerns the strength of the magnetic field inside intraspines, where convection
takes place. \citet{scharmer2006gap} and \citet{spruit2006gap} originally proposed that they would be field-free, 
thereby coining the term \emph{field-free gap}. However, most observational evidence points towards a magnetic field 
strength of at least 1 kG \citep{borrero2008pen,borrero2010pen,puschmann2010pen,tiwari2013decon,tiwari2015decon}. 
Three-dimensional magnetohydrodynamic simulations of penumbral fine structure also yield magnetic field values
of the order to 1-1.5 kG inside penumbral intraspines \citep{rempel2012mhd} irrespective of the boundary conditions and
grid resolution. \citet{spruit2010pen} interpreted the striations seen perpendicular to the penumbral filaments
in high-resolution continuum images as a consequence of fluting instability, and established 
an upper limit of $B \le 300$ Gauss for the magnetic field inside intraspines. This redefines \emph{field-free} 
to mean instead \emph{dynamically weak} magnetic fields, where
the magnetic pressure is smaller than the kinematic pressure. We note however that this interpretation has been challenged 
by \citet{barthi2012pen}, who argued that the same striations can be produced by the sideways swaying motions of the 
intraspines even if these harbor strong magnetic fields ($B \ge 1000$ Gauss).

The limited observational evidence in favor of strong convective motions perpendicular to the penumbral 
filaments, and almost complete lack of evidence for weak magnetic fields in penumbral intraspines has 
been traditionally ascribed to: {\bf (a)} the insufficient spatial resolution of the spectropolarimetric 
observations \citep[see Sect.~3.2][]{scharmer2012pen};  {\bf (b)} the smearing effects of straylight that 
are incorrectly dealt with by two-component inversions employing variable filling factors 
\citep[see Sect.~2.2 in][]{scharmer2013pen}; and {\bf (c)} the impossibility to probe layers located deep 
enough to detect them \citep[see Sect.~5.4 in][]{spruit2010pen}. In this work we will address these issues 
by employing spectropolarimetric observations of the Fe \textsc{i} spectral lines at 1565 nm recorded with the
GRIS instrument at the GREGOR Telescope. The spatial resolution is comparable to that of the Hinode/SP instrument 
and 2.5 times better than previous investigations carried out with these spectral lines. In addition, the lines 
observed by GRIS are much more sensitive to magnetic fields at the continuum-forming layer (i.e. $\log\tau_5=0$) 
than their counterparts at 630 nm. Finally, we will account for the straylight within the instrument by performing a 
deconvolution of the observations employing principal component analysis (PCA) and different point spread functions 
(PSFs). We expect that with these new data and analysis techniques we will be able to settle, in either direction, 
the dispute about the strength of the magnetic field in penumbral intraspines (e.g. filaments). A study of the convective 
velocity field will be presented elsewhere.

\section{Observations}%
\label{section:observations}

The observations employed in this work were taken with the 1.5-meter GREGOR 
telescope \citep{schmidt2012gregor} located at the Spanish Observatory of El Teide.
Our targets were two active regions: NOAA 12045 and the leading spot in NOAA 12049. They were 
observed on April 24th, 2014 between UT 9:56 and 10:10, and on May 3rd, 2014 between 
UT 14:05 and 14:26, respectively.\\

\begin{table*}
\begin{center}
\caption{Atomic parameters of the observed spectral lines. $\lambda_{0}$ is the central laboratory wavelength 
of each spectral line. $\sigma$ and $\alpha$ represent the cross-section (in units of Bohr's radius squared $a_0^2$) and 
velocity parameter of the atom undergoing the transition, respectively, for collisions with neutral atoms under the 
ABO theory \citep{abo1,abo2,abo3}.\label{table:atomicdata}}
\begin{tabular}{cccccccc}
Ion & $\lambda_{0}$\tablefootmark{a} & $\chi_{\rm low}$\tablefootmark{a} & $\log(gf)$ & Elec.conf\tablefootmark{a} & $\sigma$ & $\alpha$ & $g_{\rm eff}$ \\
 & [{\AA}] & [eV] & & & & & \\
\hline
Fe \textsc{i} & 15648.515 & 5.426 & $-$0.669\tablefootmark{b} & ${^7}D_{1}-{^7}D_{1}$ & 975\tablefootmark{b} & 0.229\tablefootmark{b} & 3.0\\
Fe \textsc{i} & 15652.874 & 6.246 & $-$0.095\tablefootmark{b} & ${^7}D_{5}-{^6}D_{4.5}4f[3.5]^{0}$ & 1427\tablefootmark{b} & 0.330\tablefootmark{b} & 1.45\\
Fe \textsc{i} & 15662.018 & 5.830 & 0.190\tablefootmark{c} & ${^5}F_{5}-{^5}F_{4}$ & 1197\tablefootmark{c} & 0.240\tablefootmark{c} & 1.5\\
\hline
\end{tabular}
\tablefoot{\tablefootmark{a}{Values taken from \citet{nave1994}}. \tablefootmark{b}{Values taken from \citet{borrero2003atomic}}.
\tablefootmark{c}{Values taken from \citet{shaun2007atomic}}}
\end{center}
\end{table*}

GREGOR's Infrared Spectrograph \citep[GRIS;][]{collados2012gregor} coupled to the Tenerife Infrared Polarimeter 
\citep[TIP2; ][]{collados2007tip} was used to record the Stokes vector $\ve{I}^{\rm obs}(\lambda)=(I,Q,U,V)$ across a 4 nm 
wide wavelength region around 1565 nm and with a wavelength sampling of $\delta_\lambda \approx 40$ m{\AA}~pixel$^{-1}$. 
This wavelength region was therefore sampled with about 1000 spectral points, out of which we selected 
a 2.4 nm wide region with $N_\lambda=600$ spectral points that includes three Fe \textsc{i} spectral lines 
(see Table~\ref{table:atomicdata}). The large Land\'e factors and wavelengths of these lines 
ensure a high sensitivity to the magnetic field. In addition, the sensitivity 
of these spectral lines to the different physical parameters, in particular the magnetic 
field strength, peaks at an optical depth five times larger than in the case 
of the widely used Fe \textsc{i} lines at 630 nm. This is a consequence the H$^{-}$ opacity having 
a minimum at 1640 nm \citep{chandra1946hminus}, which makes the Sun more transparent at 
these wavelengths compared to the visible range, and due to the large excitation potential 
of these spectral lines, which require large temperatures (i.e. deep photospheric layers) to 
populate the energy levels involved in the electronic transition. More details will be provided
in Section~\ref{subsection:depth}.\\

The effective Land\'e factors in Table~\ref{table:atomicdata} have been obtained under the 
assumption of LS coupling. This is valid for the Fe \textsc{i} spectral lines at 1564.8 nm and 1566.2 nm. 
The spectral line Fe \textsc{i} at 1565.2 nm is better described under JK coupling. However, following
\cite{bellot2000umb} we consider it to be a normal Zeeman triplet with an effective Land\'e factor of 
$g_{\rm eff}=1.45$.\\

\begin{figure}[h]
\begin{center}
\includegraphics[width=8cm]{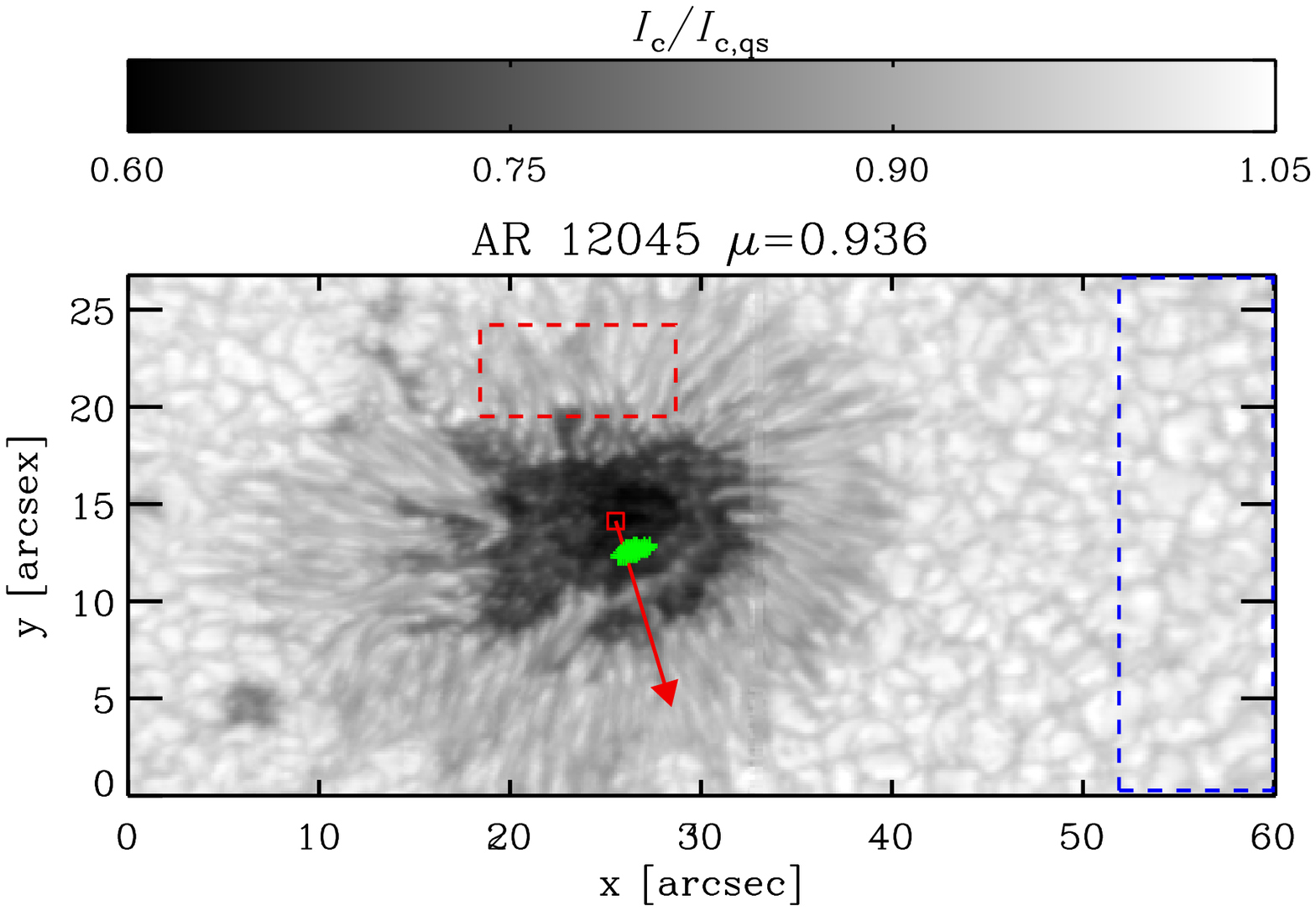} \\
\includegraphics[width=8cm]{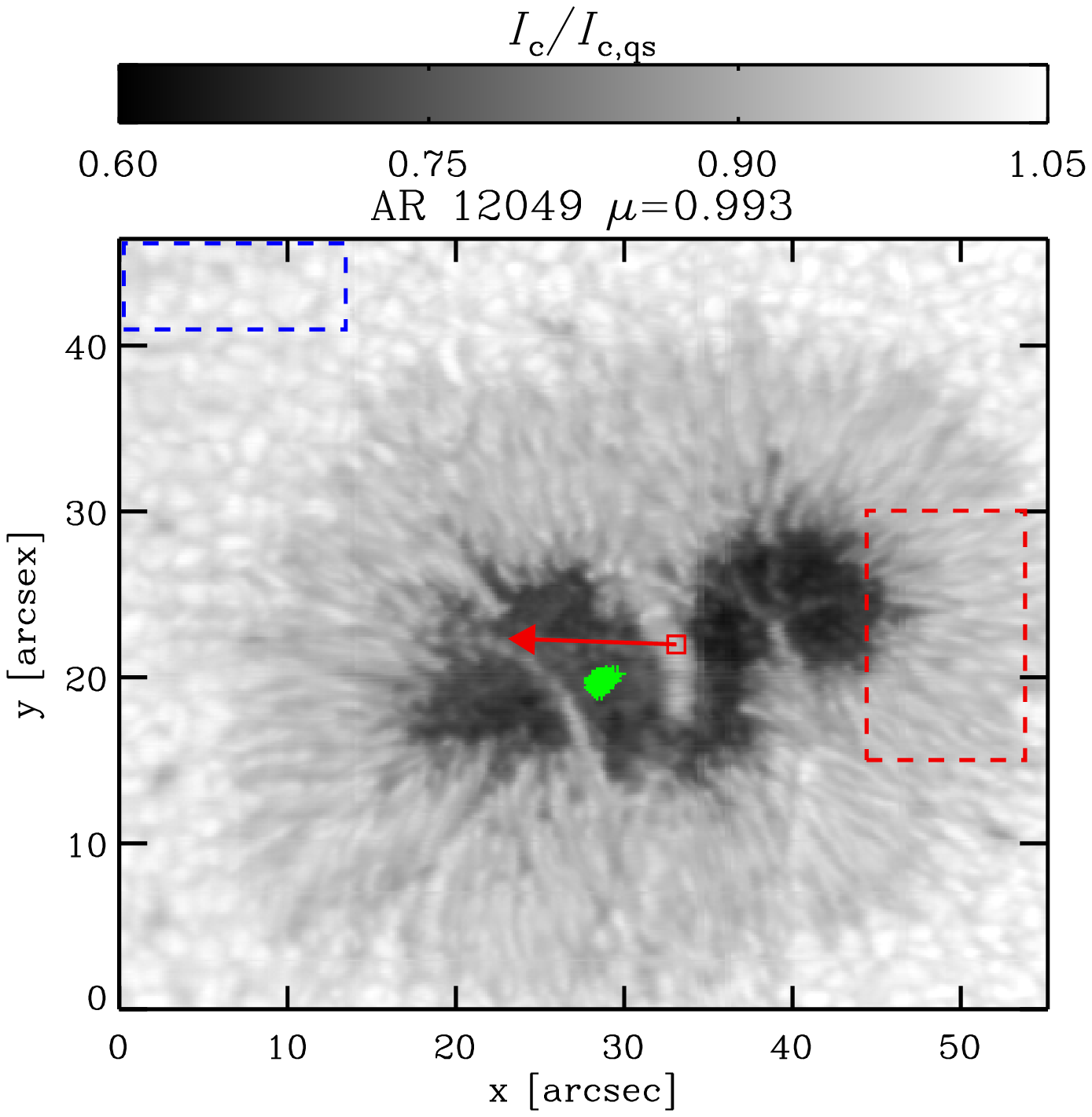}
\caption{Continuum intensity $I_{\rm c}$ (at 1565 nm) of NOAA 12045 (top) and leading spot of NOAA 12049 (bottom) normalized to the quiet 
Sun continuum ($I_{\rm c,qs}$) as observed with the GREGOR Infrared Spectrograph (GRIS) on April 24th, 2014 and May 3rd, 2015, 
respectively. The red squares denote the center of each sunspot: $(x,y)=(125\arcsec,-309\arcsec)$ for NOAA 12045 and  
$(x,y)=(73\arcsec,-83\arcsec)$ for NOAA 12049. The red arrow points towards the center of the solar disk. The areas enclosed 
by the blue-dashed rectangles have been used for calibration purposes (e.g. to calculate the normalization factor $I_{\rm c,qs}$; 
see also Sect.~\ref{section:spectralpsf}), while the areas enclosed by the red-dashed rectangles indicate the regions that 
have been analyzed in our work (see Sects.~\ref{subsection:inversion} and \ref{section:results}). The scanning direction 
(i.e. direction of movement of the spectrograph's slit) was from up to bottom along the $y$-axis. \label{figure:ic}}
\end{center}
\end{figure}

During the observing time the solar image rotated, as a consequence of GREGOR's altitude-azimuthal
mount \citep{volkmer2012gregor}, by about 5.6$^{\circ}$ and 14.7$^{\circ}$ for NOAA 12045 and NOAA 12049, 
respectively. This was sufficiently small not to
heavily distort the continuum images reconstructed from the individual slit positions. 
The GREGOR Adaptive Optics System \citep[GAOS;][]{berkefeld2012gregor} worked throughout 
the entire scans (see Sections~\ref{subsection:pnsn} and ~\ref{subsection:pcadecon}). We note that
the Fried parameter $r_0$ scales as $\lambda^{6/5}$ and therefore the size of the isoplanatic
patches is about 2.5 times larger at 1565 nm than at 630 nm, thereby allowing the AO to perform
much better (e.g. compensaring seeing and optical aberrations) at larger wavelengths. Normalized continuum 
intensity images from the observed active regions are presented in Figure~\ref{figure:ic}. The contrast of 
the continuum intensity in the granulation in these images (at 1565 nm) is about 2.2 \%. This value is equivalent
to a 5.5 \% contrast at 630 nm, which is lower than the 7.2 \% contrast seeing by Hinode/SP, thus indicating that
the spatial resolution is slightly worse than that of Hinode/SP (i.e. 0.32\arcsec). To estimate it more accurately
we have determined a cutoff frecuency between 2 and 3 arcsec$^{-1}$ in the power spectra of the continuum intensity
in the granulation, yielding a spatial resolution of about 0.4-0.45\arcsec.

By correlating our images with simultaneous HMI/SDO full-disk continuum images, we estimate that 
the sunspots' centers were located at coordinates $(x,y) = (125\arcsec,-309\arcsec)$ and $(x,y) = (73\arcsec,-83\arcsec)$ 
(measured from disk center), for NOAA 12045 and NOAA 12049 respectively. These values correspond to
heliocentric angles of $\Theta=20.5^{\circ}$ ($\mu=0.936$) and $\Theta=6.5^{\circ}$ ($\mu=0.993$). The image 
scale is also estimated by correlating the images with HMI data, yielding $\delta_x=0.135$\arcsec~pixel$^{-1}$ 
and $\delta_y=0.136$\arcsec~pixels$^{-1}$ along the $x$ and $y$ axis, respectively. The sizes of the scanned 
regions are 60$\times27$ arcsec$^2$ (top panel in Fig.~\ref{figure:ic}) and $53\times45$ arcsec$^2$ (bottom 
panel in Fig.~\ref{figure:ic}). The width of the spectrograph's slit was set to 0.27\arcsec~ (i.e. twice the 
scanning step).\\

The data have been treated with dark current substraction, flat-field correction, fringe removal \citep[see][for more details]{mortenpenumbra}
as well as polarimetrically calibrated (see Collados et al.; {\it in preparation}), yielding a noise level of about $\sigma_q \approx \sigma_u \approx 
\sigma_v \approx 10^{-3}$ in units of the quiet Sun continuum intensity. These values were achieved with 30 ms exposures per 
accumulation and a total of five accumulations per modulation step. The wavelength calibration of the data was 
performed under the assumption that the averaged quiet-Sun intensity spectral line profiles, obtained as the mean 
$I(\lambda)$ inside the blue rectangles in Fig.~\ref{figure:ic}, are located at the central laboratory wavelength 
position $\lambda_{0}$ (see Table~\ref{table:atomicdata}) but shifted by $-$535 m\,s$^{-1}$ to account for the convective 
blueshift in the observed spectral lines as determined from the Fourier Transform Spectrometer \citep[FTS;][]{livingston1991fts}. 
After the aforementioned standard calibrations, the data were corrected for spectral scattered light within the spectrograph 
(Section~\ref{section:spectralpsf}), and for the smearing effects introduced by the telescope PSF (Section~\ref{subsection:spatialpsf}).\\

\section{Spectral profile: veil correction}%
\label{section:spectralpsf}

Prior to the analysis of the recorded data we have estimated GRIS's spectral profile employing 
a similar procedure as the one described in \citet{bianda1998,allende2004fts,cabrera2007fts}. We first 
obtained a GRIS-\emph{simulated} average quiet-Sun intensity profile $I_{\rm qs}^{\rm sim}(\lambda)$
by convolving FTS data with a wavelength profile $t(\lambda)$ that mimics the effects of GRIS's spectral profile:

\begin{eqnarray}
I_{\rm qs}^{\rm sim}(\lambda) = I_{\rm fts}(\lambda) \Conv t(\lambda) \;.
\label{equation:conv}
\end{eqnarray}

\begin{figure}
\begin{center}
\includegraphics[width=8cm]{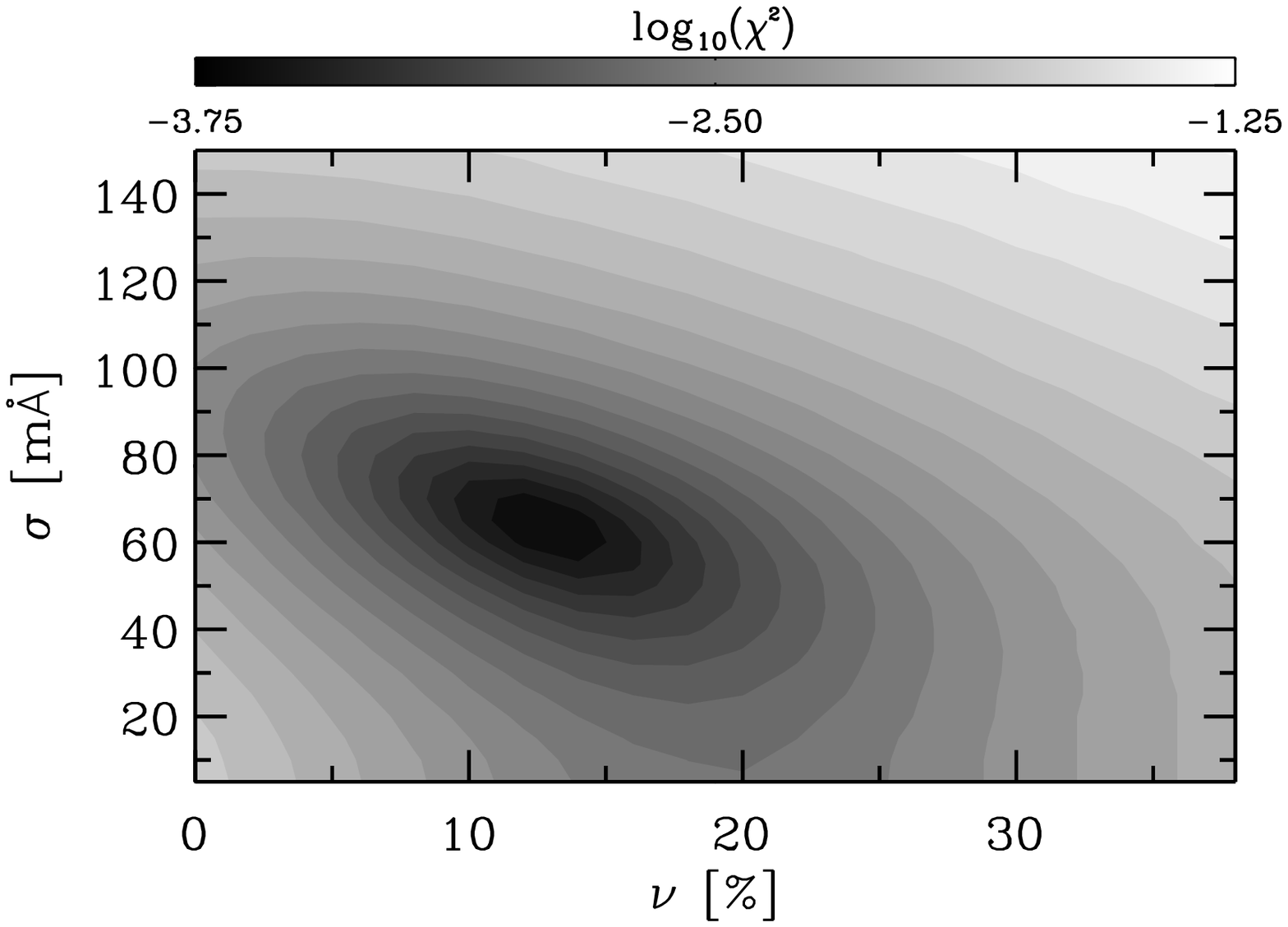} \\
\includegraphics[width=8cm]{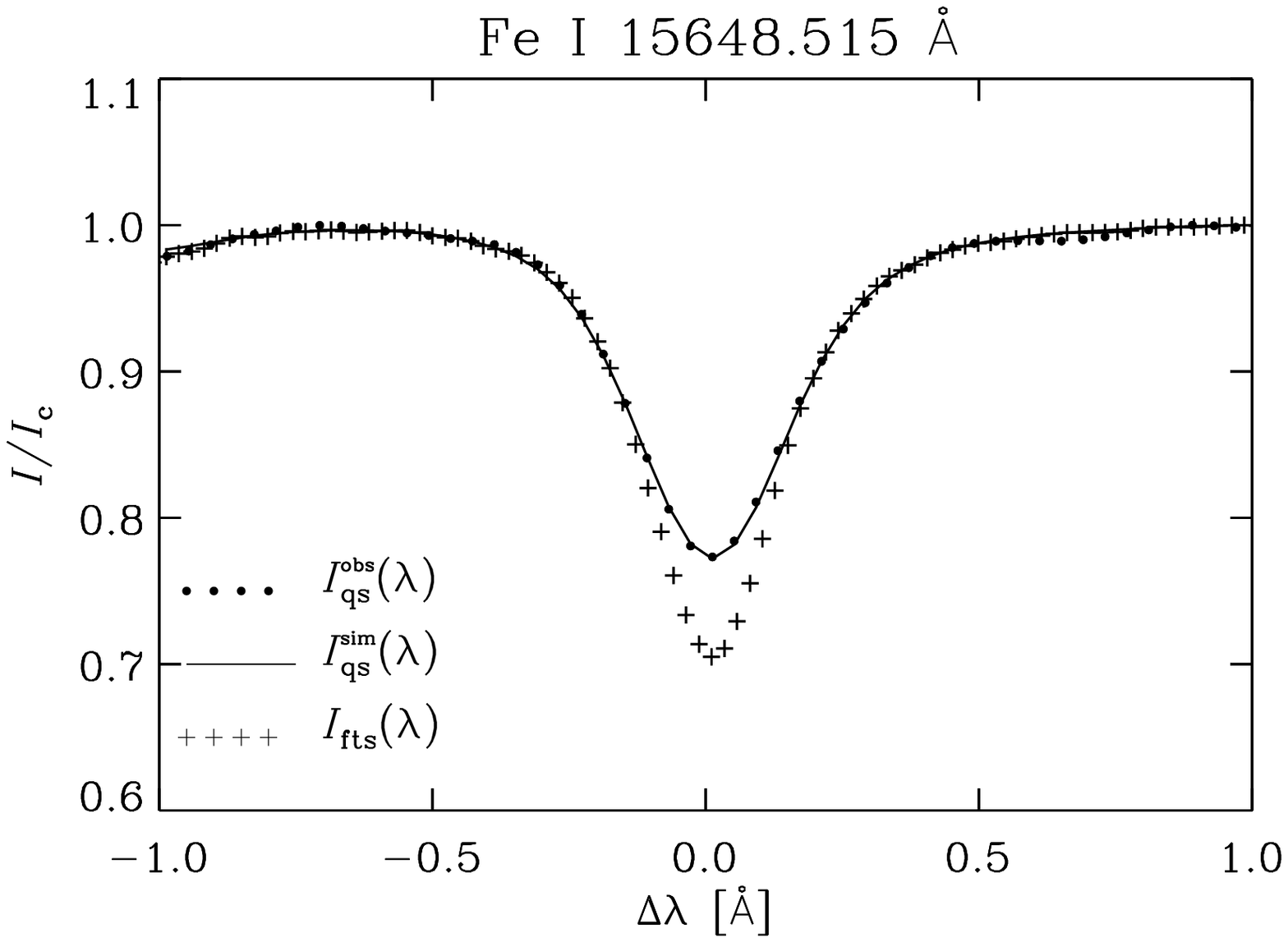}
\caption{{\it Top panel}: $\chi^2$-surface as a function of the parameters $\nu$ and $\sigma$ defining the
spectrograph's profile (Eq.~\ref{equation:veil}). This surface was obtained from the comparison 
between the observed ($I_{\rm qs}^{\rm obs}$) and simulated ($I_{\rm qs}^{\rm sim}$) quiet-Sun intensity profile of the 
Fe \textsc{i} line at 1564.8 nm. {\it Bottom panel}: FTS intensity profile of Fe \textsc{i} 1564.8 nm 
$I_{\rm fts}$ (crosses), simulated average quiet-Sun intensity profile obtained after applying the spectrograp's 
profile to FTS data $I_{\rm qs}^{\rm sim}$ (solid line), observed average quiet-Sun intensity profile 
from GRIS data $I_{\rm qs}^{\rm obs}$ in NOAA 12049 (filled circles; see also blue-dashed rectangles in Fig.~\ref{figure:ic}).
\label{figure:spectralpsf}}
\end{center}
\end{figure}

In principle $t(\lambda)$ can be approximated by a Gaussian function $g(\lambda,\sigma)$, where $\sigma$
refers to the width of spectral profile. However, a Gaussian profile decays rapidly with wavelength,
thereby neglecting the possible effect of extended wings in the spectral profile. These wings can be interpreted as 
\emph{spectral scattered-light} or \emph{spectral veil}, that mixes information coming from far-away wavelengths. 
As a first approximation one can consider this spectral veil to be wavelength
independent and proportional to the continuum intensity, in which case Equation~\ref{equation:conv} becomes:

\begin{eqnarray}
I_{\rm qs}^{\rm sim}(\lambda,\sigma,\nu) = (1-\nu) I_{\rm fts}(\lambda) \Conv g(\lambda,\sigma) + \nu I_{\rm c,fts} \;,
\label{equation:veil}
\end{eqnarray}

\noindent where $I_{\rm c,fts}=I_{\rm fts}(\lambda_c)$ is the continuum intensity in FTS data and $\nu$ is the fraction of
spectral scattered-light. Next, we create an array of simulated averaged quiet-Sun intensity profiles through 
Eq.~\ref{equation:veil}, $I_{\rm qs}^{\rm sim}(\lambda)$, employing different values of $\sigma$ and $\nu$. Each of 
these is then compared, via a $\chi^2$ merit-function, with the observed (i.e. by GRIS) average quiet-Sun 
intensity profile $I_{\rm qs}^{\rm obs}(\lambda)$ that is obtained by averaging the intensity profiles in the 
quiet-Sun region denoted by the blue-dashed rectangles in each map in Figure~\ref{figure:ic}:

\begin{eqnarray}
\chi^2(\sigma,\nu) = \sum\limits_{k=1}^{N_\lambda^{'}} \Big[I_{\rm qs}^{\rm sim}(\lambda_k,\sigma,\nu)-I_{\rm qs}^{\rm obs}(\lambda_k)\Big]^2 \;,
\label{equation:chi2conv}
\end{eqnarray}

\noindent where the index $k=1,...,N_\lambda^{'}$ runs for all wavelengths observed by GRIS across a particular spectral
line. We note that, since GRIS's spectral sampling ($\approx 40$ m{\AA} pixel$^{-1}$) is coarser than that of FTS ($\approx 23$ m{\AA} 
pixel$^{-1}$), $I_{\rm qs}^{\rm sim}$ must be re-interpolated to GRIS's wavelength grid before Equation~\ref{equation:chi2conv} 
can be evaluated. Figure~\ref{figure:spectralpsf} (top panel) shows the $\chi^2$ surface as a function of $\sigma$ and $\nu$ that
results from the comparison of $I_{\rm qs}^{\rm sim}(\lambda)$ and $I_{\rm qs}^{\rm obs}(\lambda)$ for the Fe \textsc{i} spectral line
located at 1564.8 nm (see Table~\ref{table:atomicdata}). The minimum of this surface is attained for $\sigma=70$ m{\AA} 
and $\nu=0.12$ (12 \% of spectral veil). This value of $\sigma$ corresponds to a FWHM for the spectral transmission of
$g(\lambda,\sigma)$ of approximately 165 m{\AA} \citep[cf.][]{mortenpenumbra}. Same values are obtained for both sunspots in Fig.~\ref{figure:ic}. 
Figure~\ref{figure:spectralpsf} (bottom panel) compares the original FTS intensity profile $I_{\rm fts}(\lambda)$ (crosses), 
the simulated quiet-Sun profile $I_{\rm qs}^{\rm sim}(\lambda)$ (solid lines) obtained through Eq.~\ref{equation:veil} with 
the aforementioned values of $\sigma$ and $\nu$, as well as GRIS's observed quiet-Sun profile $I_{\rm qs}^{\rm obs}(\lambda)$ in NOAA 12049 
(filled circles). As it can be seen $I_{\rm qs}^{\rm obs}(\lambda)$ and $I_{\rm qs}^{\rm sim}(\lambda)$ are extremely similar as guaranteed by the small
value of $\chi^2$ achieved (top panel in Figure~\ref{figure:spectralpsf}).

With this, it is now possible to substract the effect of the spectral veil from the observed intensity profiles 
at every pixel $(x,y)$ in the entire field-of-view and obtain corrected profiles, $I^{\rm cor}(x,y,\lambda)$, by simply applying:

\begin{eqnarray}
I^{\rm cor}(x,y,\lambda) = [1-\nu]^{-1}[I^{\rm obs}(x,y,\lambda)-\nu I^{\rm obs}_{\rm c}(x,y)] \;.
\label{equation:corrveil}
\end{eqnarray}

At this point a number of clarifications are in order. The first one is that the correction for spectral veil affects
only the intensity profiles since the continuum polarization is zero $Q^{\rm obs}_{\rm c}(x,y)=U^{\rm obs}_{\rm c}(x,y)=
V^{\rm obs}_{\rm c}(x,y)=0$ everywhere, making $Q^{\rm cor}=Q^{\rm obs}$ (and likewise for $U$ and $V$). In the following sections we 
will not deal anymore with $\ve{I}^{\rm obs}(x,y,\lambda)$ but instead will consider only the \emph{veil-corrected} Stokes 
vector $\ve{I}^{\rm cor}(x,y,\lambda)$. However, for the sake of simplicity we will continue to refer to it as $\ve{I}^{\rm obs}(x,y,\lambda)$.\\
Finally it must be borne in mind that by applying Eq.~\ref{equation:corrveil} here we are only
correcting for the spectral veil. The rest of the spectrograph's profile, namely the Gaussian profile
$g(\lambda,\sigma)$ in Eq.~\ref{equation:veil} (with $\sigma=70$ m{\AA}), will be considered at a later step in the 
analysis (Sect.~\ref{subsection:inversion}).\\

\section{Analysis}
\label{section:analysis}

\subsection{Spatial Point Spread Function}
\label{subsection:spatialpsf}

Seeing, scattered light and diffraction effects cause the observed Stokes vector to differ from the one emitted
at a point $(x,y)$ on the solar surface. Their effect can be quantified via the point spread function (PSF) of the
optical system $P(x,y)$ which smears out the original signal:

\begin{equation}
\ve{I}^{\rm obs}(x,y,\lambda) = \iint \ve{I}^{\rm sun}(\xp,\yp,\lambda) P(x-\xp,y-\yp) \df\xp \df\yp .
\label{equation:psfconv}
\end{equation}

The above equation indicates that some percentage of the signal from points $(\xp,\yp)$ with $\xp \ne x$
and $\yp \ne y$, will contaminate the Stokes vector at the location $(x,y)$.
Cleaning the observed signal from this effect has been traditionally limited to data from space-borne
instrumentation where the optical system has a well-defined $P(x,y)$ function that can be calculated, 
sometimes even analytically, for all sorts of instruments \citep{danilovic2008psf,wedemeyer2008psf,mathew2009psf,asensio2010psf,
feller2014psf}. From the ground the situation becomes more complicated, as the time-varying seeing imposes strong limits 
as to how fast the observations must be carried out in order to be corrected. In this case $P(x,y)$ must 
be determined empirically via reconstruction techniques such a Speckle-reconstruction \citep{keller1992speckle}, 
Multi-Object Multi-Frame Blind Deconvolution \citep{lofhadl2002momfbd,michiel2005momfbd} or Phase Diversity \citep{paxman1996pd}. 
For these reasons, in ground-based observations, the aforementioned procedure has been limited to high-throughput filter-based 
spectropolarimeters \citep{scharmer2008crisp,bello2008gfpi,moro2010ibis,valentin2011imax}. While in principle these 
techniques should deliver diffraction limited observations, recently \cite{scharmer2012pen} and 
\cite{lofdahl2012stray} have argued that high-altitude seeing remains uncorrected.

Nowadays, high-order adaptive optics allow us to obtain spectropolarimetric data from slit-based instruments 
that are stable enough, during acquisition, so as to attempt to decontaminate $\ve{I}^{\rm obs}$ and retrieve 
$\ve{I}^{\rm sun}$ (Eq.~\ref{equation:psfconv}). The issue remains however as to how to obtain a PSF that represents the optical
system, including seeing, in ground-based long-slit spectrographs \citep{beck2011decon}. Although the AO-system can correct most of the low- and
mid-order optical aberrations, it cannot correct all of them. Moreover the correction weakens for regions away from the 
AO lockpoint. Consequently the observations are never completely diffraction limited and hence the difficulties in having
a good knowledge of $P(x,y)$. We will therefore resort to indirect means to obtain a meaningful PSF that can be employed
to isolate the solar signal. Let us first assume, for simplicity, that the PSF can be described by two different Gaussian functions corresponding
to narrow- "\emph{n}" and wide- "\emph{w}" angles contributions, respectively:

\begin{equation}
P(x,y) = p_n g_n(x,y,\sigma_n) + p_w g_w(x,y,\sigma_w) \;,
\label{equation:psf2gauss}
\end{equation}

\noindent where $\sigma_n$ and $\sigma_w$ correspond to the distances, in seconds of arc, in which each Gaussian contributes with 
$\approx 68.2$ \% of its power. $p_n$ and $p_w$ denote the relative contribution from narrow and wide angles respectively.
We note that $P(x,y)$, $g_w$, and $g_n$ are normalized to unity, and that $p_w+p_n=1$.

\subsubsection{Estimation of $p_w$ and $\sigma_w$}
\label{subsection:pwsw}

For sunspots not far from disk center there is always a region, located within the umbra, where
the magnetic field is aligned with the observer's line-of-sight. To locate these regions within
the observed FOV we have selected those pixels in Fig.~\ref{figure:ic}, where $I_{\rm c}/I_{\rm c,qs} < 0.7$
and where the maximum of the total linear polarization  $\textrm{max}|(Q^2+U^2)^{1/2}| < 5\times 10^{-3}$.
 In total there are about 60 pixels inside the umbra of each sunspot that fulfill these conditions. 
They are marked in Fig.~\ref{figure:ic} with green crosses. We refer to the location of these pixels as 
$(\xas,\yas)$. \\

Let us also remember that, whenever a magnetic field is aligned with the observer's line-of-sight, the intensity profile 
$I(\lambda)$ of a normal Zeeman triplet ($J_l=0 \rightarrow J_u=1$ or vice-versa) presents two (and only two) distinct 
absorption features according to the selection rule $\Delta M = \pm 1$ for the Zeeman effect \citep{deltoro2003book}. 
Each of these two absorption features is shifted with respect to the central wavelength by an amount that is proportional 
to $\propto (g_{\rm eff} B \Delta M)$, where $B$ is the modulus of the magnetic field. If the separation between these two 
components is sufficiently large, they appear as two unblended spectral lines, and therefore the observed intensity at the 
central wavelength must be very close to the continuum intensity: $I(\lambda^{\dagger}) \approx I_{\rm c} = I(\lambda_{\rm c})$. 
To ensure that these conditions are met we will use the intensity profile of Fe \textsc{i} 1564.8 nm because this is the spectral 
line in our observations with the largest Land\'e factor (see Table~\ref{table:atomicdata}) and the only one featuring a 
normal Zeeman triplet. We note that $\lambda^{\dagger}$ does not necessarily correspond to the central laboratory wavelength 
$\lambda_0$. Instead it is defined as the wavelength located half the way between the two absorption features with $\Delta M = \pm 1$.\\

\begin{figure}
\begin{center}
\includegraphics[width=8cm]{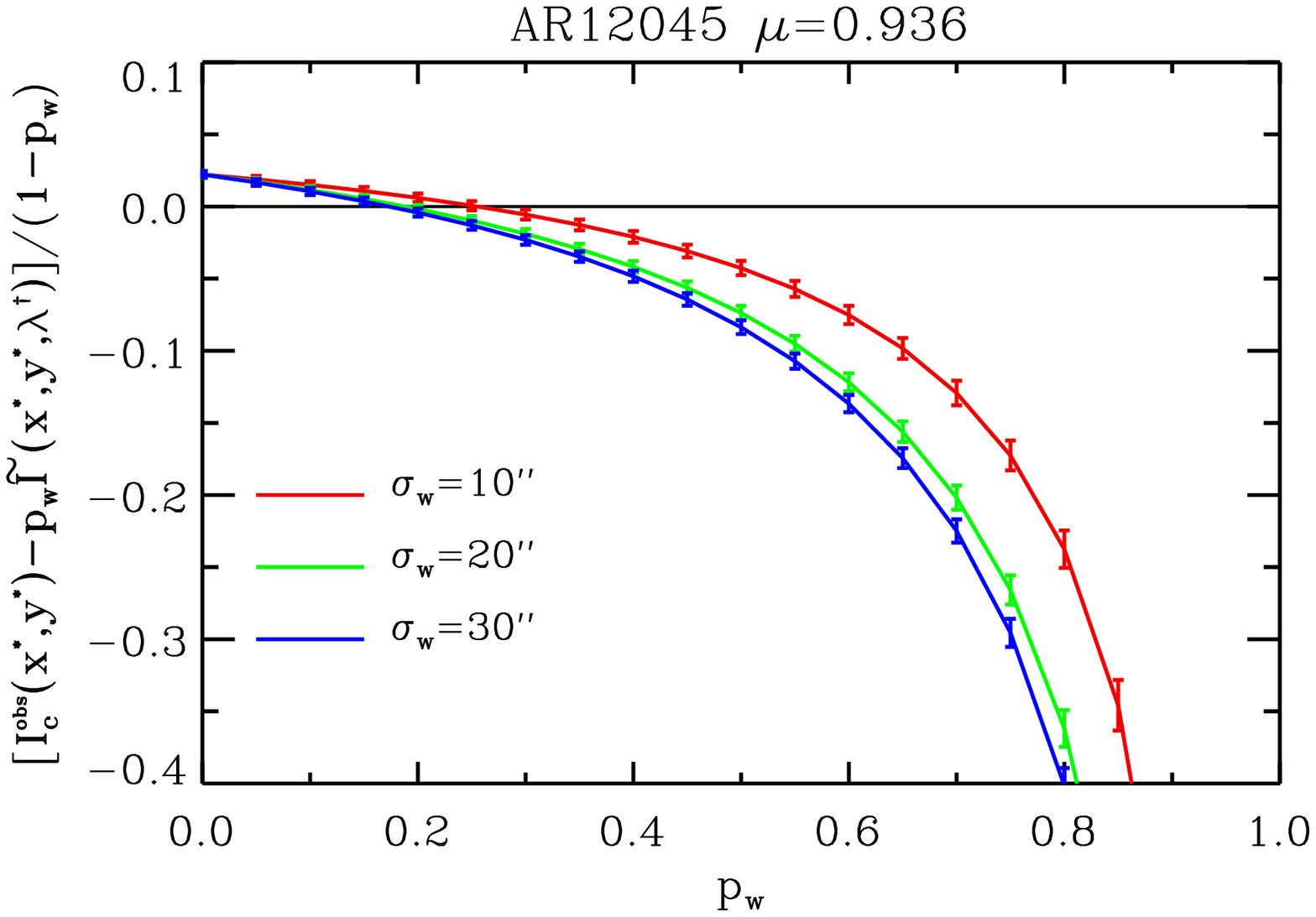} \\
\includegraphics[width=8cm]{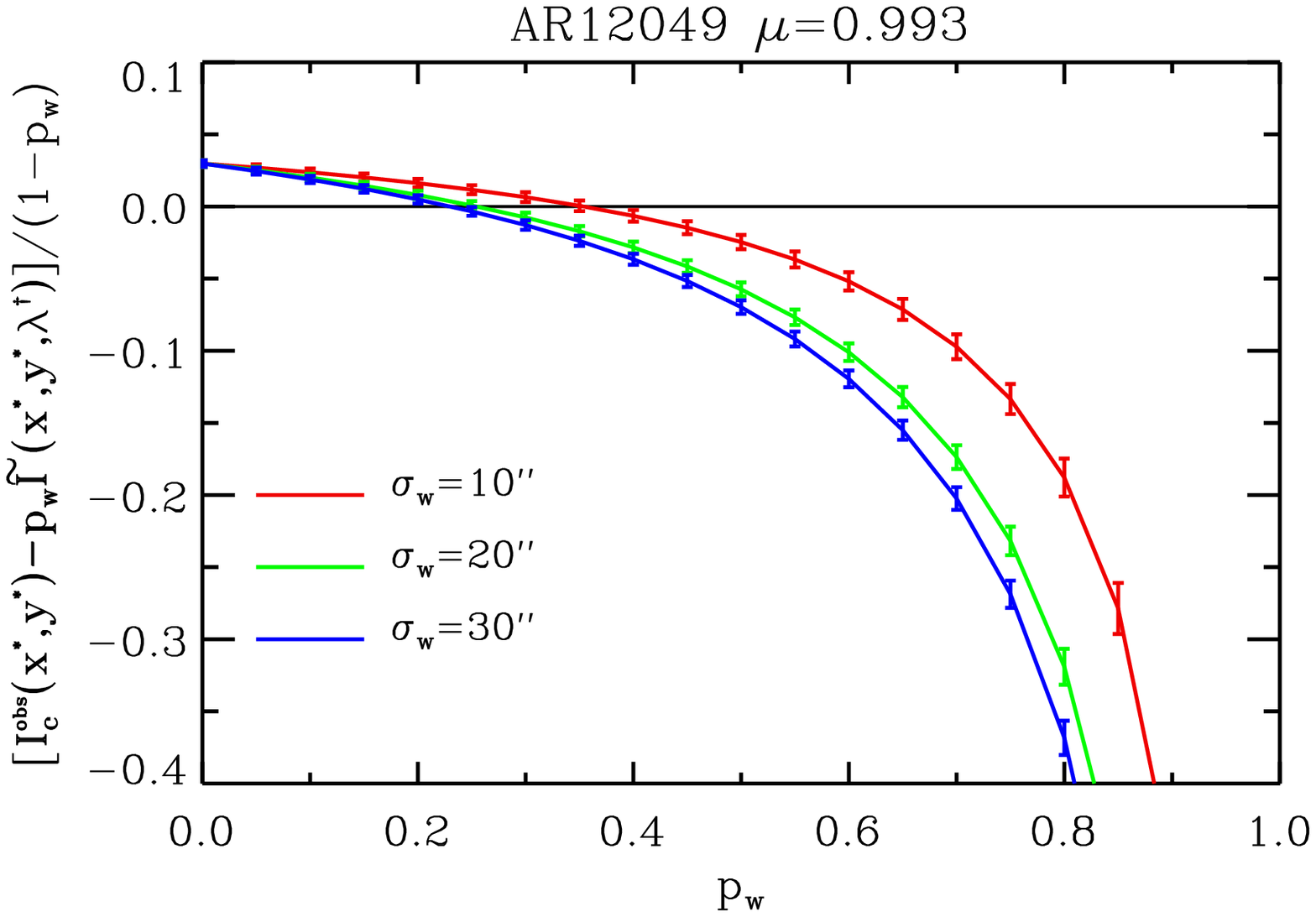}
\caption{Determination of the amount of stray light $p_w$ (wide-angle contribution) by finding the value of $p_w$
for which the intensity at central wavelength $\lambda^{\dagger}$ in those pixels $(\xas,\yas)$, where
the magnetic field is aligned with the observer's line-of-sight is equal to the continuum intensity. This
occurs for $p_w=0.2-0.3$ depending on the width $\sigma_w$ of the Gaussian employed to mimic the effects of the
straylight. The vertical error bars refer to the standard deviation in the determination of $p_w$ using each
of the pixels in $(\xas,\yas)$.\label{figure:pwsw}}
\end{center}
\end{figure}

Interestingly, the profiles selected above and located at $(\xas,\yas)$ do exhibit a small absorption feature at $\lambda^{\dagger}$.
As the magnetic field is mostly aligned to the observer's line-of-sight in those pixels, the absorption feature at $\lambda^{\dagger}$
is unlikely to be produced by the magnetic field (unshifted $\Delta M=0$ component of the Zeeman pattern). Instead, it must arise from 
an absorption profile unaffected by the magnetic field (i.e. penumbra and quiet-Sun surrounding both sunspots). In order 
to determine how far this contribution comes from and how much there is, we calculate a new intensity profile
at each of the selected pixels $\widetilde{I}(\xas,\yas)$ in the following way:

\begin{equation}
\widetilde{I}(\xas,\yas,\lambda) =  \iint g_w (x\!-\!\xas,y\!-\!\yas,\sigma_w) I^{\rm obs}(x,y,\lambda) \df x \df y 
\label{equation:intensity_sigma}
\end{equation}

\noindent where $g_w$ was defined in Eq.~\ref{equation:psf2gauss}. Next, we calculate the difference
between the continuum intensity at the location $(\xas,\yas)$, that is $I^{\rm obs}_{\rm c}(\xas,\yas)$,
and  $p_w \widetilde{I}(\xas,\yas,\lambda^{\dagger})$ for different values of $p_w$ and $\sigma_w$. Since 
the bulk of the contribution to the absorption feature observed at $\lambda^{\dagger}$ is ascribed to the
penumbra and granulation and these are located about 10\arcsec~ away from $(\xas,\yas)$, then $\sigma_w \ge 10$\arcsec~. 
The results are presented in Figure~\ref{figure:pwsw} for the two sunspots in our dataset. As it can be seen the intensity 
at the central wavelength position
becomes comparable to the continuum intensity, $I^{\rm obs}(\lambda^{\dagger}) \approx I^{\rm obs}_{\rm c}$, for $p_w \approx 0.2-0.3$.
This indicates that the absorption feature seen at $\lambda^{\dagger}$ in those pixels, where the magnetic field
is aligned with the observer's line-of-sight, $(\xas,\yas)$, can be explained with a 20-30 \% straylight contamination coming
from outside the umbra. Unfortunately the exact distance cannot be reliably determined since the same values of $p_w$ are
obtained for three different values of $\sigma_w$. As a compromise we will adopt $\sigma_w=20"$ and $p_w=0.2$. It should be noted
however that the smaller $\sigma_w$ the larger must $p_w$ be in order to explain the absorption feature seen
at $\lambda^{\dagger}$. This happens because $\widetilde{I}(\xas,\yas,\lambda)$ does not include
a strong absorption at $\lambda^{\dagger}$ for small values of $\sigma_w$, as $g_w$ (see Eq.~\ref{equation:intensity_sigma})
mainly includes contributions from the neighborhood of $(\xas,\yas)$ (i.e. sunspot umbra).\\

\begin{figure}
\begin{center}
\includegraphics[width=9cm]{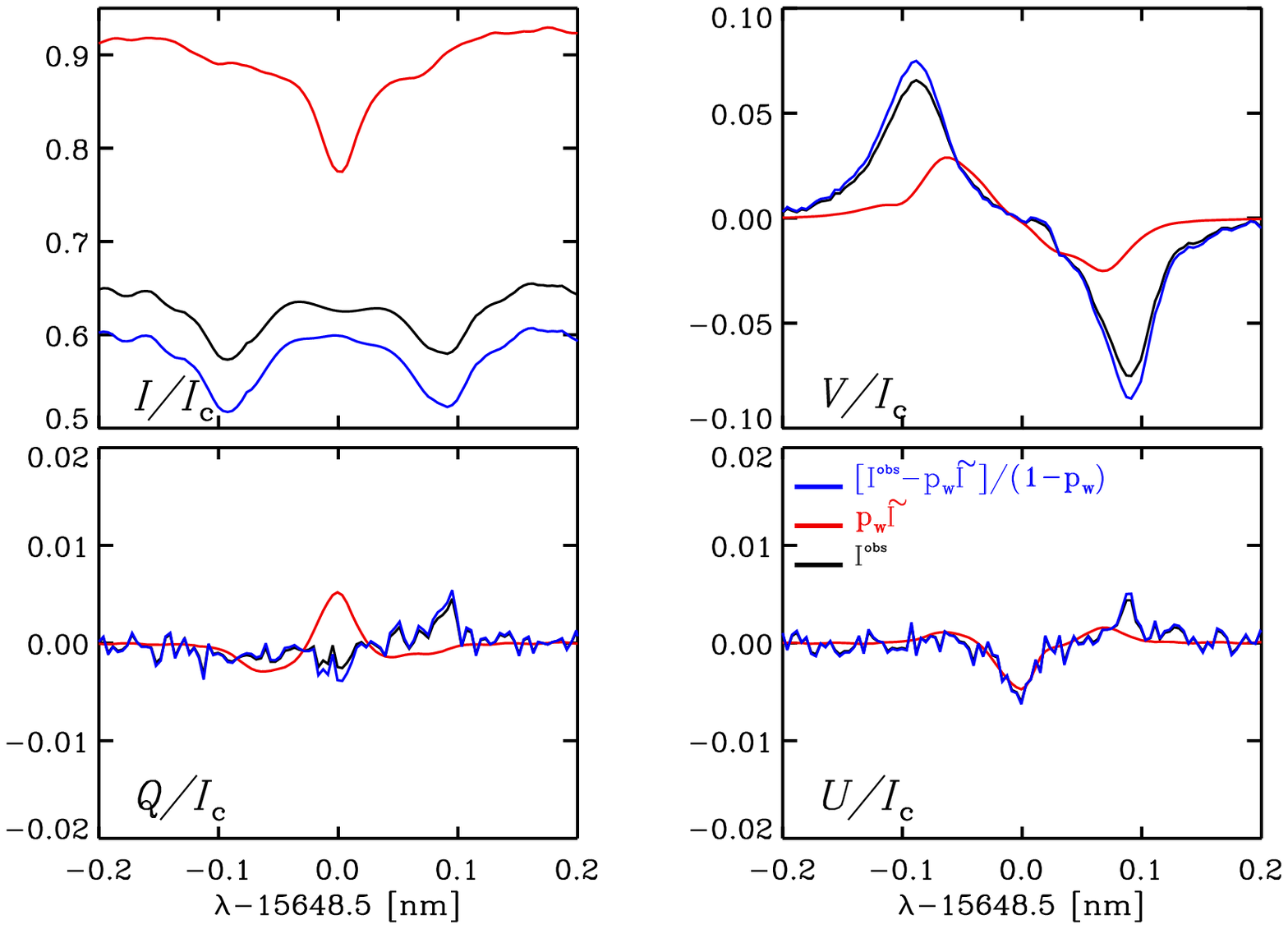} \\
\caption{Example of the determination of $\sigma_w$ and $p_w$ via the removal of the $\pi$-component in the intensity profile (Stokes $I$; 
top-left panel) in a pixel where the magnetic field is aligned with the observer's line-of-sight. Black lines represent the observed 
Stokes profiles $\ve{I}^{\rm obs}$. Red lines show the wide-angle contribution (i.e. scattered light) from the surrounding penumbra and quiet Sun,
$\tilde{\ve{I}}$. Blue lines represent the observed Stokes profiles after removing 20 \% of the wide-angle contribution. See text for details.
\label{figure:stray_example}}
\end{center}
\end{figure}

The procedure described above is illustrated, for one of the pixels referred to as $(\xas,\yas)$, in Figure~\ref{figure:stray_example},
This figure depicts the observed Stokes profiles $\ve{I}^{\rm obs}$ in black lines. As it can be seen Stokes $I$ (upper-left panel) 
has a small absorption feature at a wavelength $\lambda^{\dagger} \approx 1564.85 $ [nm] located in between the two $\sigma$-components 
of the Zeeman pattern. The very small observed $Q$ (bottom-left) and $U$ (bottom-right) profiles, along with the large circular 
polarization signals (Stokes $V$; top-right), implies the magnetic field in this pixel is mostly aligned with the observer's line-of-sight.
In red lines we show the contribution from the surrounding penumbra and quiet-Sun $\tilde{\ve{I}}$ as calculated 
through Eq.~\ref{equation:intensity_sigma} assuming that about 68 \% of the wide-angle scattered light comes from a distance, around 
the considered pixel, of 20\arcsec: $\sigma_w = 20\arcsec$. Blue lines show the umbral profile in the same pixel after removing 
$p_w \tilde{\ve{I}}$ from $\ve{I}^{\rm obs}$ using $p_w=0.2$ and renormalizing using $(1-p_w)$. The resulting Stokes $I$ profile
(blue lines in top-left panel) does no longer feature the absorption at $\lambda^{\dagger}$ (i.e. where the unshifted $\Delta M=0$ would appear). 
Indeed, at this wavelength the observed intensity is the same as the continuum intensity.\\

It is important to mention at this point that the absorption feature seen at $\lambda^{\dagger}$ could also be caused by unudentified 
molecular blends that are common in the umbra at near-IR wavelengths, left-over fringes, and even by the magnetic field not being fully aligned 
with the observer's line-of-sight. It could also appear as a combination of all the above. Becuase we have assumed only one origin, the values 
for $p_w$ obtained in this section should be considered only as an upper limit of the real amount of wide-angle scattered light present in our 
observations.

\subsubsection{Estimation of $p_n$ and $\sigma_n$}
\label{subsection:pnsn}

Once $p_w$ is known, it is straightforward to determine $p_n$ as $p_n=1-p_w=0.8$. The width of the narrow-angle
PSF, $\sigma_n$, is determined by performing a slit-scan of the pinhole array while it is inserted
into the light-path at the third focal point along the optical path (i.e. before the spectrograph's slit). By 
fitting the shape of the light curve at the pinhole discontinuity, a value of 3.2 pixels on the CCD was determined.
Considering the values for the image scale given in Sect.~\ref{section:observations} this yields $\sigma_n=0.18"$,
corresponding to a FWHM of 0.43" and very similar to the spatial resolution estimated from the power spectrum
of the granulation (Sect.~\ref{section:observations}). This value is larger than the theoretical diffraction-limited 
FHWM of $\approx 0.27"$\footnote{This number includes only the primary mirror. A slightly larger value is obtained if the cetral obscuration and spiders
from the secondary mirror are included.}. While we have not investigated in detail the reason for this worse-than-ideal
performace we can point to a number of possible sources such as high-altitude seeing \citep{scharmer2012pen,lofdahl2012stray},
width of the spectrograph's slit, and slightly off-focus spectrograph.

\subsection{PCA expansion and deconvolution}
\label{subsection:pcadecon}

Now that we have empirically estimated the PSF of the optical system $P(x,y)$ we can attempt to retrieve
$\ve{I}^{\rm sun}$ by deconvolving $\ve{I}^{\rm obs}$ through Eq.~\ref{equation:psfconv}. One possibility would be to deconvolve
individual two-dimensional monochromatic images (e.g. $I^{\rm obs}(x,y,\lambda_k)$ for $k=1,...,N_\lambda=600$ and likewise for $Q$,
$U$, and $V$). However, this approach neglects the information contained in the wavelength dependence of the Stokes parameters. 
This is important because this information allows to reduce the influence of the noise in the deconvolution process. 
Currently there are two methods that take advantage of this wavelength dependence. The first one,
referred to as \emph{spatially-coupled inversions} \citep{vannoort2012decon}, uses the radiative transfer equation for polarized
light to exploit the wavelength dependence of the Stokes vector during the deconvolution process \citep{tino2013decon,vannoort2013decon,
lagg2014decon,tiwari2015decon}. This method has the advantage that is physically driven, but it requires deep modifications
in existing inversion codes for the radiative transfer equation. The second method is based on the Principal Component
Analysis (PCA) of the data \citep{basilio2013pen,quintero2015pca}. Despite being statistically driven, it has the advantage that it can be
used in combination with any existing inversion codes for the radiative transfer equation without further modifications. Because
of its simplicity, we have chosen the second of the aforementioned methods for our analysis. It must be borne in mind, however,
that it remains to be proven that both methods give the same results when applied to the same data.

Hence we follow \citet{basilio2013pen} and \citet{quintero2015pca} and expand, each of the four components 
of the Stokes vector ($\ve{I}^{\rm obs}$ and $\ve{I}^{\rm sun}$) over the entire field of view 
in Figure~\ref{figure:ic}, in a set of orthonormal eigen-vectors $\phi(\lambda)$ such that:\\

\begin{eqnarray}
\begin{split}
I^{\rm obs}_{m}(x,y,\lambda) & = \sum\limits_{n=1}^{N_\lambda} \C_{m,n}^{\rm obs}(x,y)\phi_{m,n}(\lambda) \simeq \\ &
 \simeq \sum\limits_{n=1}^{N_m^{\rm pca}} \C_{m,n}^{\rm obs}(x,y)\phi_{m,n}(\lambda) \; .
\end{split}
\label{equation:pcaexpobs}
\end{eqnarray}

\begin{eqnarray}
\begin{split}
I^{\rm sun}_{m}(x,y,\lambda) & = \sum\limits_{n=1}^{N_\lambda} \C_{m,n}^{\rm sun}(x,y)\phi_{m,n}(\lambda) \simeq \\ &
\simeq \sum\limits_{n=1}^{N_m^{\rm pca}} \C_{m,n}^{\rm sun}(x,y)\phi_{m,n}(\lambda) \; ,
\end{split}
\label{equation:pcaexpsun}
\end{eqnarray}

\noindent where the index $m=1,..,4$ corresponds to any of the four components of the Stokes vector, $I$, ..., $V$,
respectively, and the index $n=1,...,N_\lambda$ runs for the total number of observed wavelengths 
(see Sect.~\ref{section:observations}) on the upper part of the 
above equations. The summation over index $n$ is truncated to $N_{m}^{\rm pca}$, hence the approximate symbol, in the 
lower part of the equations. This will be explained later. We note the implicit assumption that the same set of 
$n$-eigenvectors $\phi_{m,n}(\lambda)$ can be used to expand the $m$-th component of both the observed and solar 
Stokes vector. The PCA analysis provides a method to calculate the eigenvectors $\phi$ and 
eigenvalues $\C$ by diagonalizing (i.e, via the Singular Value Decomposition method) the matrix of the observed 
Stokes profiles \citep{quintero2015pca} or, equivalently, the correlation matrix of the observed Stokes profiles 
\citep{skumanich2002pca,casini2013pca}. Once these have been obtained, we substitute Eqs.~\ref{equation:pcaexpobs} and ~\ref{equation:pcaexpsun} 
into Eq.~\ref{equation:psfconv}:

\begin{eqnarray}
\begin{split}
& \sum\limits_{n=1}^{N_\lambda} \C^{\rm obs}_{m,n}(x,y) \phi_{m,n}(\lambda)= \\ & \sum\limits_{n=1}^{N_\lambda} \Big[ \iint \C^{\rm sun}_{m,n}(\xp,\yp)
P(x-\xp,y-\yp) \df\xp \df\yp \Big] \phi_{m,n}(\lambda) \;.
\end{split}
\label{equation:convexp}
\end{eqnarray}

Since $\phi_{m,n}$ are orthogonal, Eq.~\ref{equation:convexp} must hold independently for each of the $N_\lambda$ eigenvectors:

\begin{eqnarray}
\C^{\rm obs}_{m,n}(x,y)\!=\!\iint \C^{\rm sun}_{m,n}(\xp,\yp) P(x-\xp,y-\yp) \df\xp \df\yp \;,
\label{equation:convcoef}
\end{eqnarray}

\noindent which shows that the original problem of convolution/deconvolution of the Stokes vector
(Eq.~\ref{equation:psfconv}) has been narrowed down to determining the coefficients $\C_{m,n}(x,y)$ 
of the expansion. Since the number of coefficients is equal to the number of observed wavelengths 
for all four Stokes parameters, $m\times n = 4 N_\lambda$, applying Eq.~\ref{equation:convcoef}
or Eq.~\ref{equation:psfconv} is, in principle, equivalent and requires the same effort. Interestingly,
only a small number of eigenvectors provide useful information about the Stokes profiles. This 
implies that we can truncate the expansions in Eqs.~\ref{equation:pcaexpobs}
and Eqs.~\ref{equation:pcaexpsun} (lower part of these equations) to a much smaller number of coefficients 
$n=1,...,N_m^{\rm pca} \|| N_{\lambda}$. The truncation provides an approximation to the $m$-th component of the 
Stokes vector $I^{\rm pca}_m(\lambda)$ (hereafter referred to as PCA-reconstructed Stokes profile) that differs 
by an amount $\mathcal{O}_m(\lambda)$ from the observed one $I_m^{\rm obs}(\lambda)$:

\begin{equation}
\mathcal{O}_m(x,y,\lambda) = \|I_m^{\rm pca}(x,y\lambda)-I_m^{\rm obs}(x,y,\lambda)\|  \;.
\label{equation:pcadif}
\end{equation}

The above equation provides a tool to determine where the expansion (Eqs.~\ref{equation:pcaexpobs},~\ref{equation:pcaexpsun})
must be truncated or, equivalently, a way to determine $N_m^{\rm pca}$. This is done by adding new eigenvectors until the mean
(spatial and spectral) difference between the PCA-reconstructed and observed Stokes profile is at the level of the noise in
the observations:

\begin{equation}
\sigma_m^2 \approx (N_x N_y N_\lambda)^{-2} \sum\limits_{i,j,k} \mathcal{O}_{m}(x_i,y_j,\lambda_k)^2 \;,
\label{equation:pcaerr}
\end{equation}

\noindent where $N_x$ and $N_y$ are the total number of spatial points along the $x$ and $y$ directions, respectively.
$\sigma_m$ refers to the noise in the $m$-th component of the Stokes vector. All these values have been provided in, or can be
obtained from, Sect.~\ref{section:observations}. Table~\ref{table:pcacoef} presents the optimum values of $N_m^{\rm pca}$
obtained through the application of the two above equations. We note that the number of coefficients needed to properly
reproduce the observed Stokes vector is larger for NOAA 12045 than for NOAA 12049. The reason for this is that NOAA 12049 is
very close to disk center ($\Theta=6.5^{\circ}$) and therefore there is very little difference between the observed Stokes
profiles on the center-side and limb-side of the penumbra. At larger heliocentric angles (NOAA 12045; $\Theta=20.5^{\circ}$) 
this is not the case anymore. Moreover, for the latter sunspot the limb-side penumbra displays highly asymmetric three-lobe
Stokes profiles similar to those in \citet{borrero2004pen,borrero2005pen}. This is undoubtedly a sign of two distinct polarities 
present in the resolution element and it explains the larger number of coefficients needed to reproduce $\ve{I}^{\rm obs}$
in NOAA 12045.

\begin{table}
\begin{center}
\caption{Number of PCA coefficients $N_m^{\rm pca}$ needed to reproduce the observed Stokes profiles ($m=1$ for $I$, $m=2$ for
$Q$, $m=3$ for $U$, and $m=4$ for $V$) at the level of the noise through Eqs.~\ref{equation:pcadif} and \ref{equation:pcaerr}.}
\begin{tabular}{c|cccc}
Active region & $N_1^{\rm pca}$ & $N_2^{\rm pca}$ & $N_3^{\rm pca}$ & $N_4^{\rm pca}$ \\
\hline
NOAA 12045 & 8 & 10 & 10 & 10 \\
NOAA 12049 & 5 & 5 & 5 & 5 \\
\hline
\end{tabular}
\label{table:pcacoef}
\end{center}
\end{table}

\begin{figure*}
\begin{center}
\begin{tabular}{cc}
\includegraphics[width=8cm]{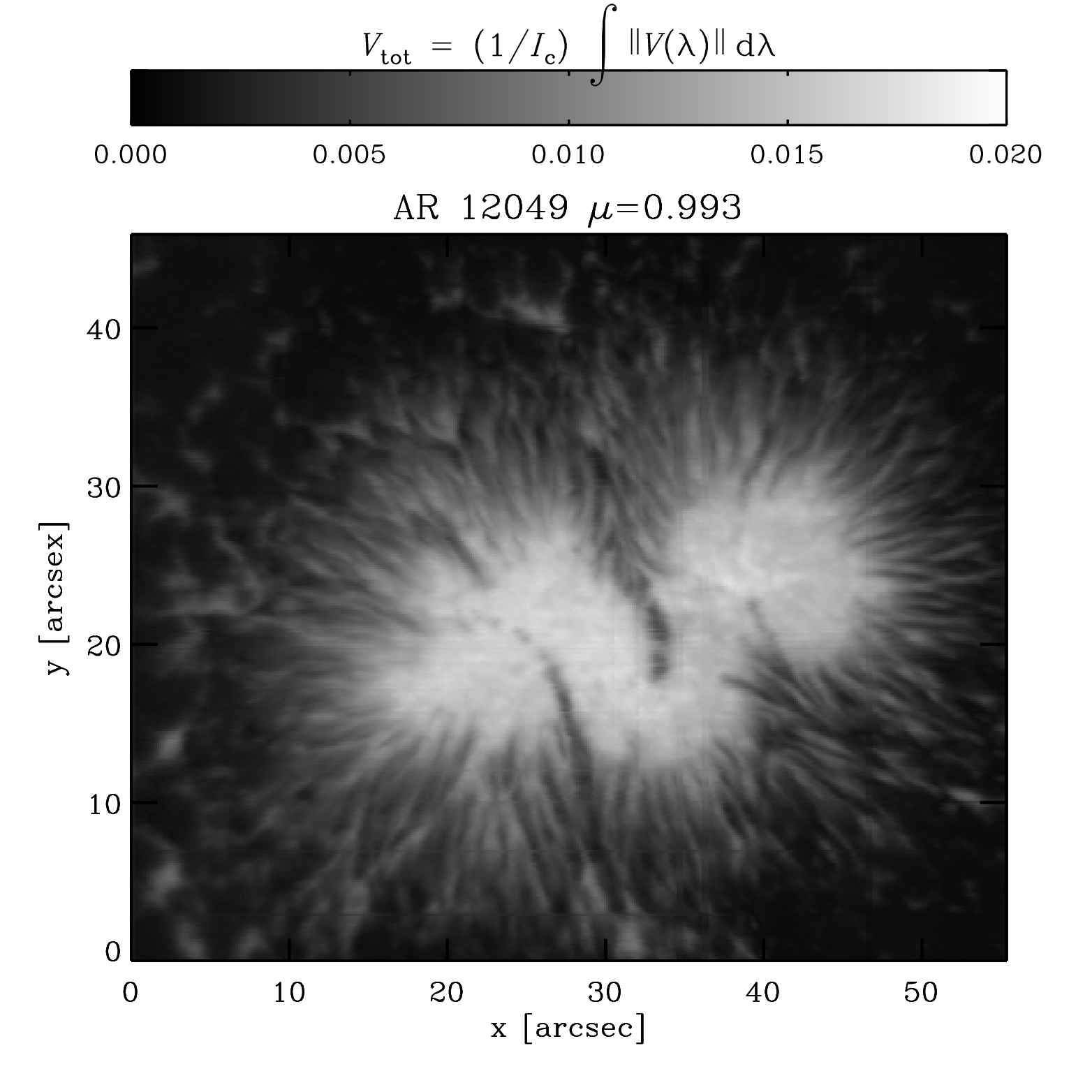} &
\includegraphics[width=8cm]{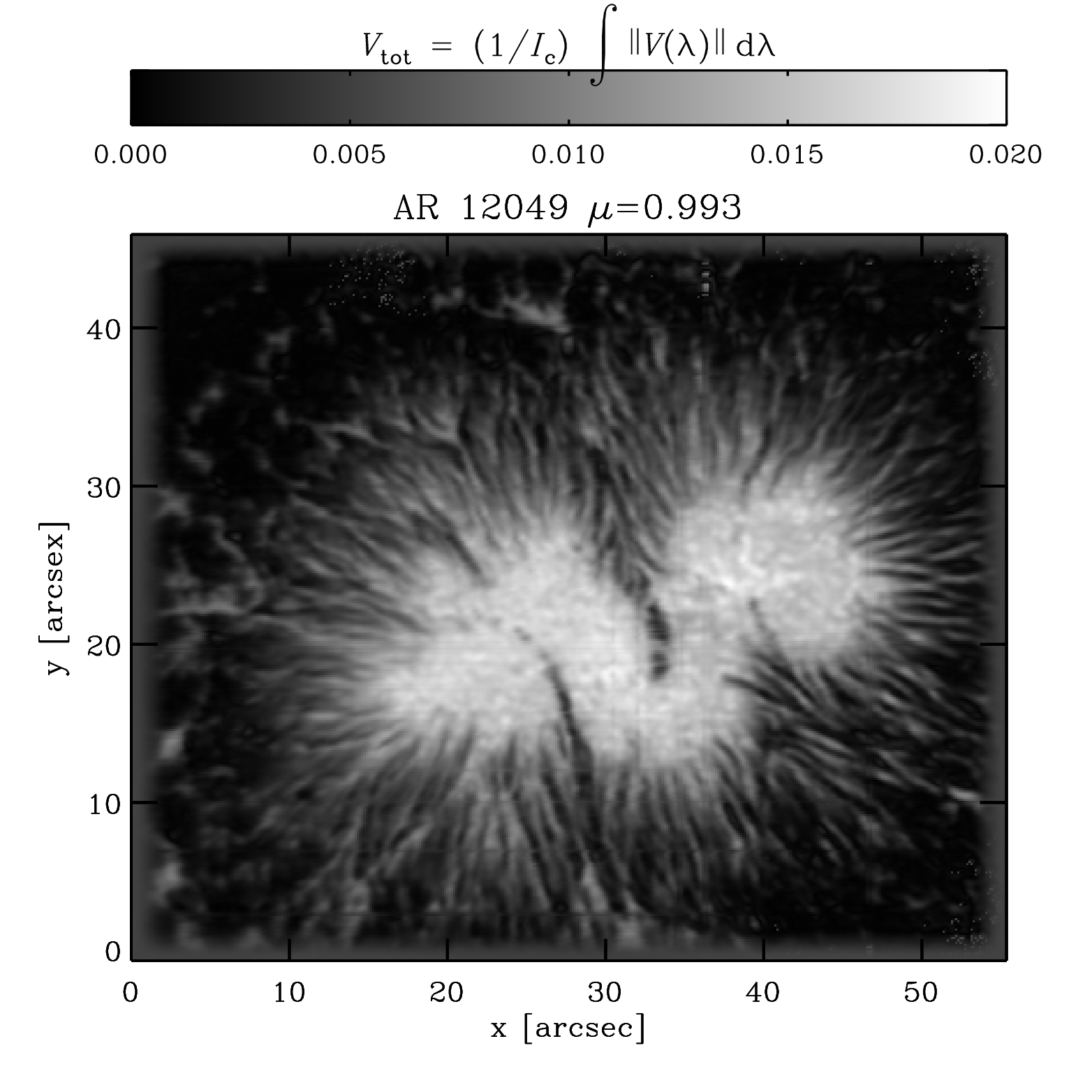} \\
\includegraphics[width=8cm]{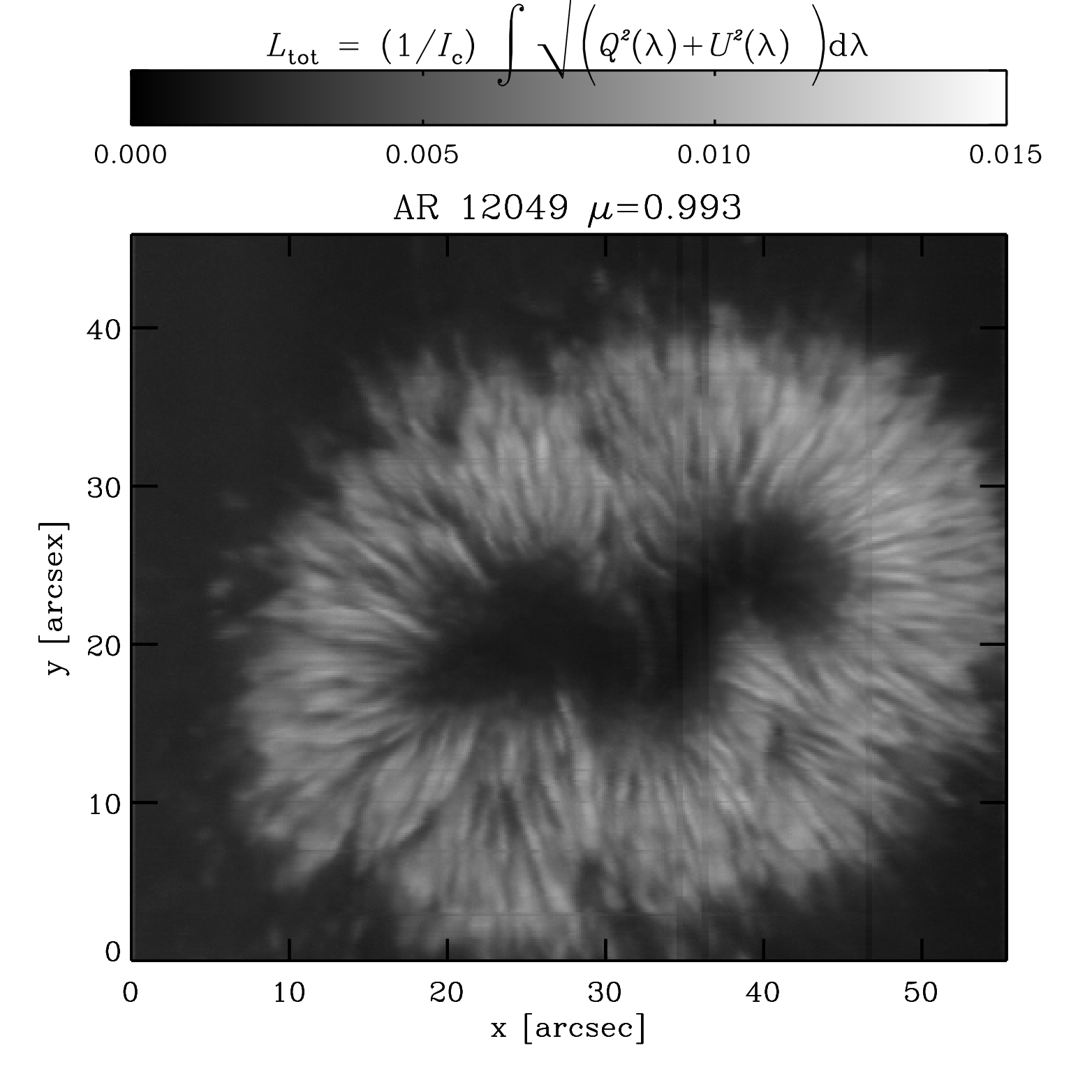} &
\includegraphics[width=8cm]{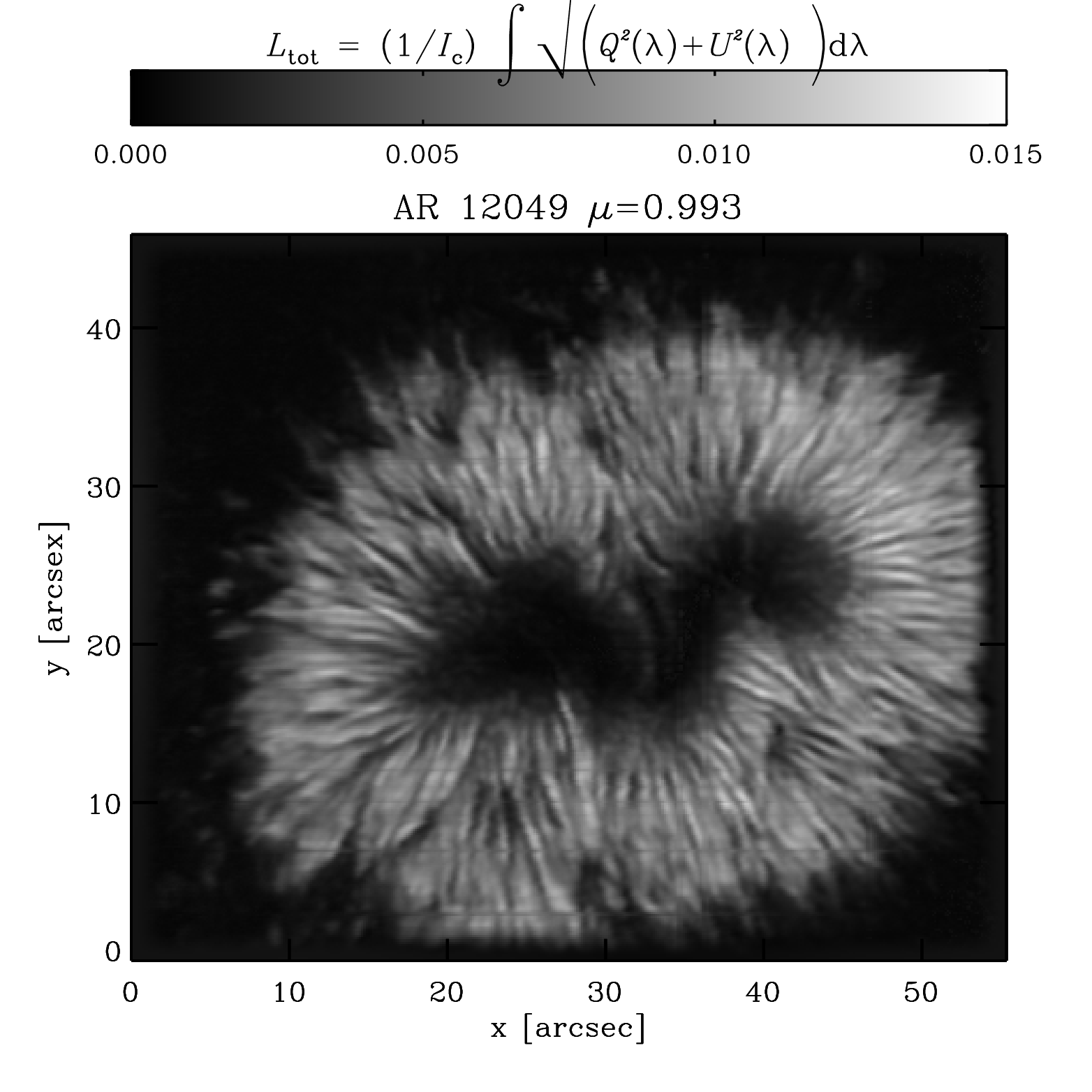} \\
\end{tabular}
\caption{Comparison between the total circular (top) and linear (bottom) polarization using the original observed Stokes
vector $\ve{I}^{\rm obs}$ (left) and the deconvolved Stokes vector $\ve{I}^{\rm sun}$ (right) after truncating the PCA expansion.
The frame around the deconvolved maps (better seen on the upper-right panel) appears as a consequence of apodization 
before deconvolution.
\label{figure:deconpca}}
\end{center}
\end{figure*}

The deconvolution of the PCA coefficients $\C_{m,n}$ for $m=1,...,4$ and $n=1,...,N_{m}^{\rm pca}$, that is obtaining
$\C_{m,n}^{\rm sun}$ from $\C_{m,n}^{\rm obs}$ and the known PSF through Eq.~\ref{equation:convcoef}, is done by
applying 10 iterations of a Lucy-Richardson-like algorithm \citep{richardson1972decon,lucy1974decon} while apodizing
the data on the 5 \% outermost region of the observed field-of-view for each sunspot. More details about this procedure
can be found in \citet{quintero2015pca}. After the $\C_{m,n}^{\rm sun}$ have been obtained, $I_m^{\rm sun}$ can be reconstructed 
via Eq.~\ref{equation:pcaexpsun}, but truncating the summation at $N_m^{\rm pca}$ from Table~\ref{table:pcacoef} instead 
of at $N_\lambda$. This yields $\ve{I}^{\rm sun}(x,y,\lambda)$. As a demonstration of the deconvolution process we present, 
in Figure~\ref{figure:deconpca}, a comparison between the total circular (top panels) and linear polarization 
(lower panels), $V_{\rm tot}$ and $L_{\rm tot}$, in NOAA 12049 obtained from the originally observed Stokes profiles 
$\ve{I}^{\rm obs}$ (left) and the deconvolved Stokes profiles $\ve{I}^{\rm sun}$ using a truncated PCA expansion 
($\simeq$ symbols in Eqs.~\ref{equation:pcaexpobs},~\ref{equation:pcaexpsun}). $V_{\rm tot}$ and $L_{\rm tot}$ are 
obtained as the wavelength integral of $\|V(\lambda)\|$ and $\sqrt\{Q^2(\lambda)+U^2(\lambda)\}$, respectively.
At high frequencies (narrow-angle) data is irredeemable lost and the the deconvolution process cannot recover it. This is not the
case of low frequencies (wide-angle scattered light), where the information can be efficiently recovered. Therefore, we can consider 
that, after the deconvolution, the spatial resolution is given by the width of the narrow-angle Gaussian: $\sigma_n=0.18"$ 
(Sect.~\ref{subsection:pnsn}). This value is also supported by the power spectra of the granulation (see Sect.~\ref{section:observations}).

\subsection{Inversion of Stokes profiles}%
\label{subsection:inversion}

We now apply the SIR inversion code \citep[Stokes Inversion based on Response Functions; ][]{basilio1992sir} to
$\ve{I}^{\rm sun}(\lambda)$ in order to infer the kinematic, thermodynamics 
and magnetic properties of the solar atmosphere in the region within by the red rectangles in Fig.~\ref{figure:ic}.
The inversion code employs an initial guess model for the solar atmosphere as a function of the continuum optical-depth
at a reference wavelength of 500 nm, $\log\tau_5$, referred to as $\ve{M}_0(\log\tau_5)$, to solve the radiative transfer 
equation and obtain a synthetic Stokes vector $\ve{I}^{\rm syn}(\lambda)$. This synthetic Stokes vector is then compared with the real one 
$\ve{I}^{\rm sun}(\lambda)$ via a $\chi^2$-merit function. The initial model is then perturbed, 
$\ve{M}_1=\ve{M}_0+\delta\ve{M}$ so as to minimize the $\chi^2$-merit-function. To that end, the perturbation 
$\delta\ve{M}$ is obtained via a Levenberg-Marquardt algorithm \citep{press1986num} and 
Singular Value Decomposition method \citep{golub1965svd}. The perturbations are introduced at specific 
$\log\tau_5$ positions called \emph{nodes}, with the final optical-depth dependence $\delta\ve{M}(\log\tau_5)$ being
obtained by interpolating between the nodes. The perturbative process is repeated until a minimum in $\chi^2$ is reached. 
The resulting atmospheric model could be taken as representative of the physical conditions of the solar plasma, however,
due to the dependence of the results from the initial guess model, we have repeated the inversion procedure ten different 
times employing random initial models $\ve{M}_0$. The one that yields the best $\chi^2$ is the one adopted as 
the final solution.

The process described above must be applied independently to each $(x,y)$ pixel contained in the red rectangles in 
Fig.~\ref{figure:ic}. Once this is done we obtain $\ve{M}(x,y,\log\tau_5)$, which consists of the three-dimensional
structure of the line-of-sight-velocity $v_{\rm los}$, temperature $T$, magnetic field strength $B$, inclination of 
the magnetic field with respect to the observer's line of sight $\gamma$, and finally the azimuth of the magnetic field
on the plane perpendicular to the observer's line-of-sight $\psi$.

Our inversions were carried out with three nodes in $T$ and two nodes in $v_{\rm los}$, $B$, $\gamma$, and $\psi$,
respectively. This adds up to a total of 10 free parameters that are used to fit at each spatial pixel $(x,y)$,
the solar Stokes vector $\ve{I}^{\rm sun}(\lambda)$ containing $4N_\lambda=2400$ data points. The node selection
will stay the same until Section~\ref{subsection:nodedepen}, where a more complex situation
will be considered. When constructing the $\chi^2$-merit-function that measures the difference between the synthetic
$\ve{I}^{\rm syn}(\lambda)$ and solar $\ve{I}^{\rm sun}(\lambda)$ Stokes vector, the polarization profiles $Q$, $U$
and $V$ were given twice the statistical weight of $I$. This is done because Stokes $I$ is more affected
by unidentified molecular blends and left-over fringes that are not fully corrected during data reduction (see
Sect.~\ref{section:observations}). We note that at each iteration step
and before being compared with $\ve{I}^{\rm sun}$, the synthetic Stokes vector $\ve{I}^{\rm syn}$ was convolved
with a Gaussian profile with $\sigma=70$ m{\AA}. This is done to include the effects of the spectrograph's 
profile $g(\lambda,\sigma)$ that remained unaccounted for after the removal of the spectral veil 
(see Sect.~\ref{section:spectralpsf}). We emphasize that we employ only one component during the inversion. 
No non-magnetic component is used to model the straylight because this was already accounted
for in the deconvolution (Sect.~\ref{subsection:pcadecon}).

\begin{figure}
\begin{center}
\includegraphics[width=8cm]{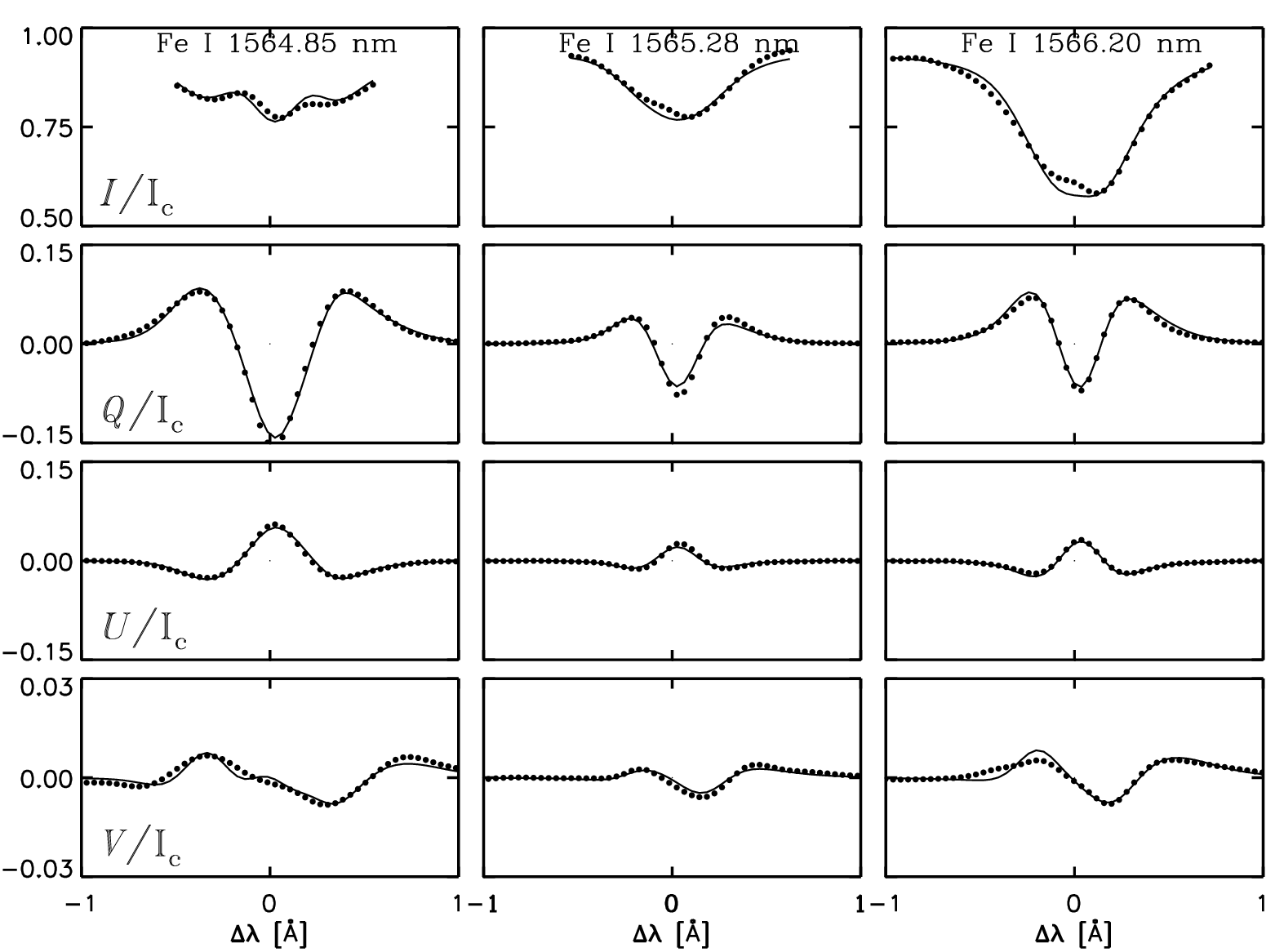}
\caption{Example of the observed (after PCA deconvolution; cilled circles) and best-fit (solid lines) 
Stokes profiles resulting from the inversion (see Sect.~\ref{subsection:inversion}) in a intrapinal
pixel. From top to bottom: results for $I$, $Q$, $U$, and $V$. From left to right: results for
Fe I 1565.85 nm, Fe I 1565.28 nm, and Fe I 1566.20 nm (see Table~\ref{table:atomicdata}). The location
of this particular pixel is indicated in Fig.~\ref{figure:mag_ar12049} with a black square.\label{figure:fitobs}}
\end{center}
\end{figure}

Figure~\ref{figure:fitobs} shows an example of the observed after PCA deconvolution (filled circles) and best-fit 
(solid lines) Stokes profiles in the three observed spectral lines. This is the result for a pixel that corresponds 
to a penumbral intraspine in NOAA 12049 (black square in Fig.~\ref{figure:mag_ar12049}). Owing to the fact
that less weight was given to the intensity during the inversion (because of blends and left-over fringes) the fits
in Stokes $I$ are always of less quality than in $Q$, $U$, $V$. Since the selected pixel is located in a intraspine,
Stokes $V$ features several lobes. Despite using only linear gradients (i.e. two nodes in $v_{\rm los}$, $B$, $\gamma$, and $\psi$), 
the inversion does a reasonable job in fitting these multi-lobed circular polarization signals. Outside the intraspines, Stokes $V$ 
regains it regular shape and the quality of the fits improves significantly.

The three dimensional distribution of the kinematic and thermodynamics parameters provided by the inversion will 
be discussed in a future paper. In this work we will focus only on the magnetic field.
The radiative transfer equation allows us to determine the magnetic field in spherical coordinates in the
observer's reference frame: $\ve{B}=(B,\gamma,\psi)$. Instead, it is more useful to analyze the magnetic field
in cylindrical coordinates in the local reference frame: $\ve{B}=(B_r,B_\theta,B_z)$, where $B_z$ corresponds to the
direction perpendicular to the solar surface, and $B_r$ and $B_\theta$ are the magnetic field components in 
polar coordinates on the plane parallel to the solar surface with origin at the sunspots' center. To this end we have employed the methods described by 
\citet{borrero2008pen} and \citet{borrero2011pen} to convert from the observer's to the local reference frame. This
method solves the 180$^{\circ}$-ambiguity in the azimuth of the magnetic field vector by choosing,
at each pixel $(x,y)$ on the solar surface, either $\psi$ or $\psi+\pi$ so that the magnetic field becomes
as radial as possible. This yields $B_r(x,y,\log\tau_5)$, $B_\theta(x,y,\log\tau_5)$ and $B_z(x,y,\log\tau_5)$.

\section{Results and discussion}%
\label{section:results}

\subsection{Field-free gaps and return flux}
\label{subsection:fieldfree}

\begin{figure*}
\begin{center}
\begin{tabular}{cc}
\includegraphics[width=8cm]{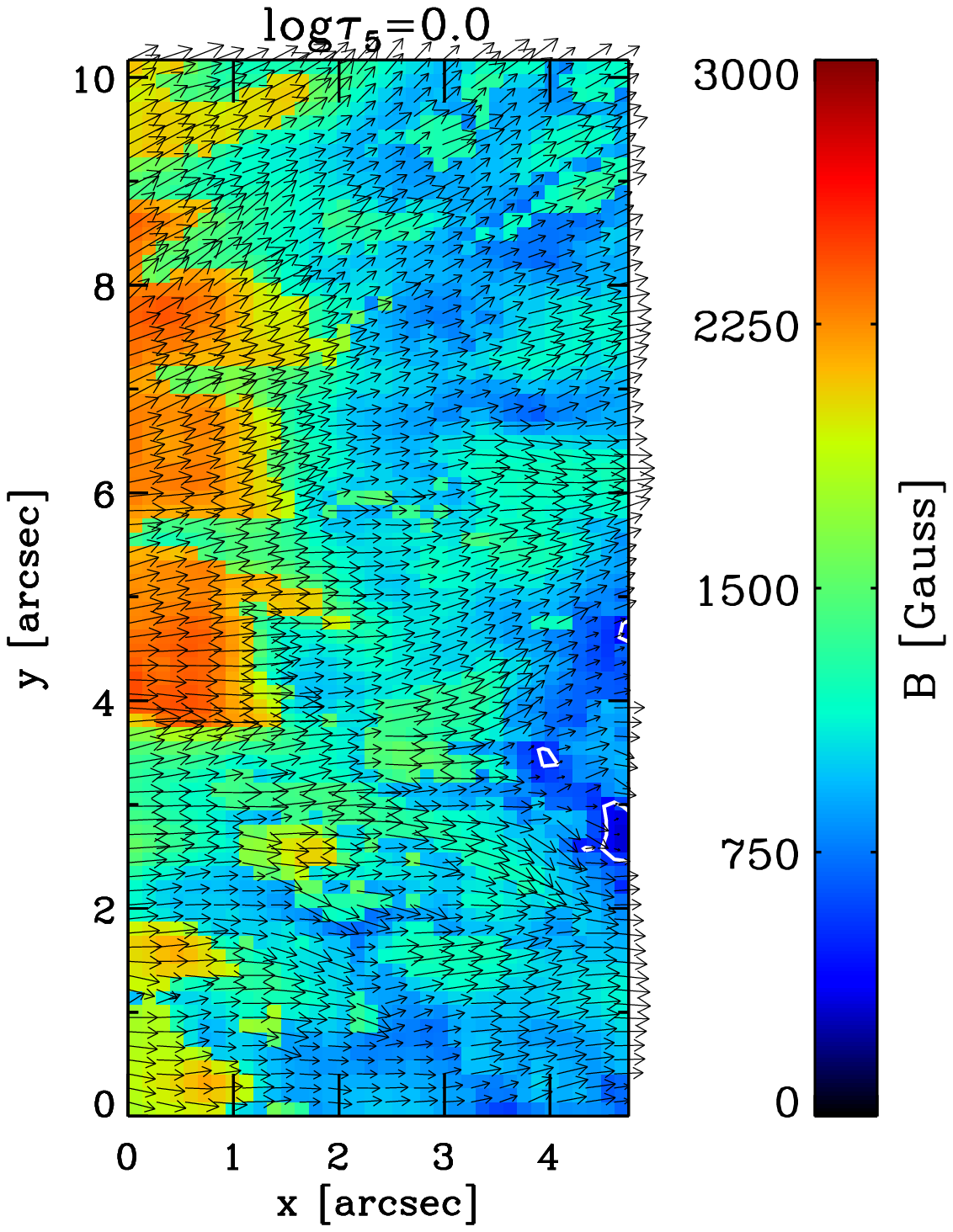} &
\includegraphics[width=8cm]{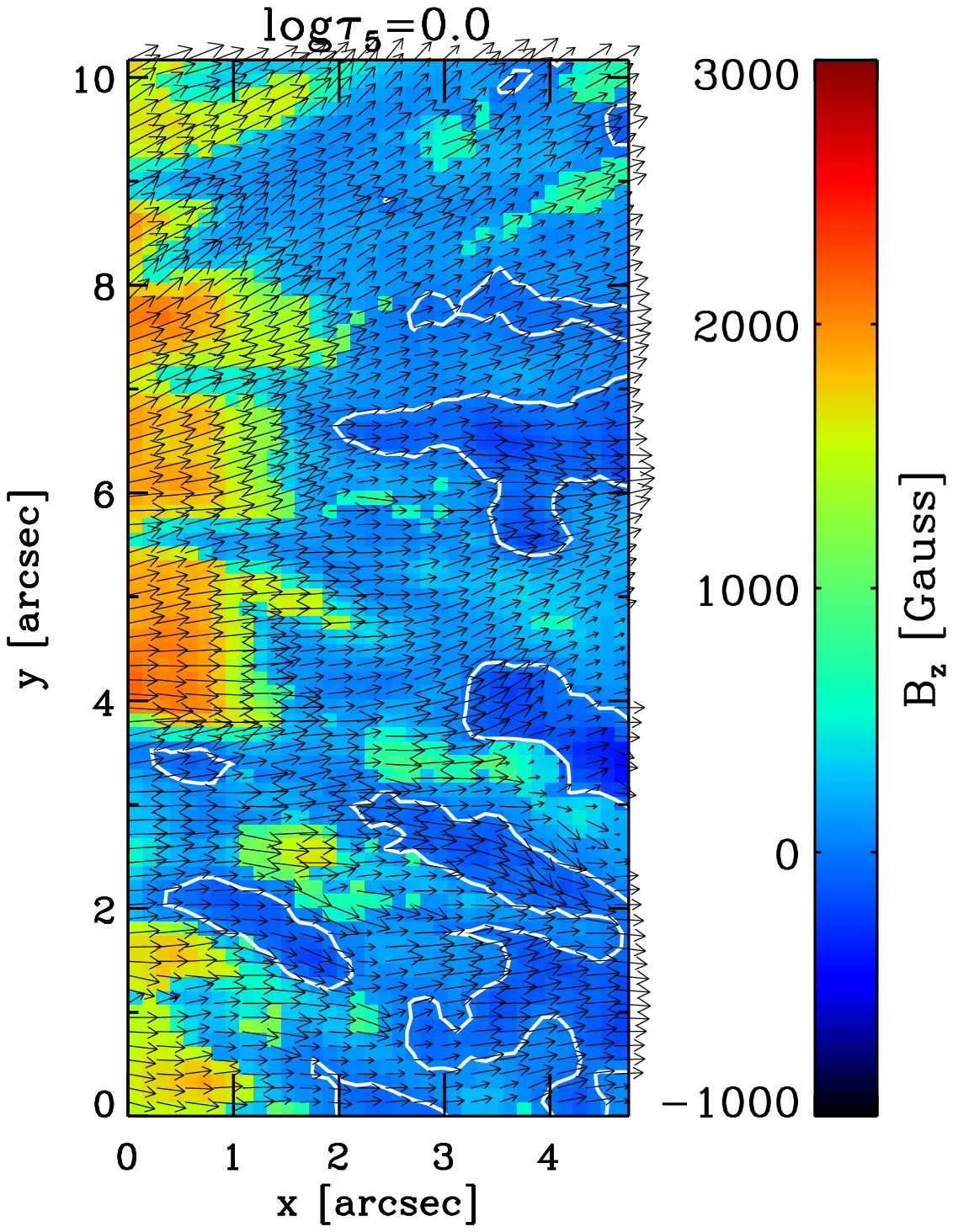} \\
\end{tabular}
\caption{Magnetic field strength $B(x,y,\log\tau_5=0)$ (left) and vertical component
of the magnetic field in the local reference frame $B_z(x,y,\log\tau_5=0)$ (right) in NOAA 12045
observed on April 24th, 2014 at $\Theta=20.5\deg$. White contours on the left and right panels 
indicate regions where $B<500$ Gauss and $B_z<0$, respectively. Black arrows indicate the projection
of the magnetic field vector $\ve{B}$ on the plane of the solar surface. This region corresponds to 
the red rectangle in Fig.~\ref{figure:ic} (top panel).\label{figure:mag_ar12045}. This map was
obtained from the inversion of the deconvolved data assuming $p_w=0.2$ (20 \% wide-angle scattered light)}
\end{center}
\end{figure*}

\begin{figure*}
\begin{center}
\begin{tabular}{cc}
\includegraphics[width=9cm]{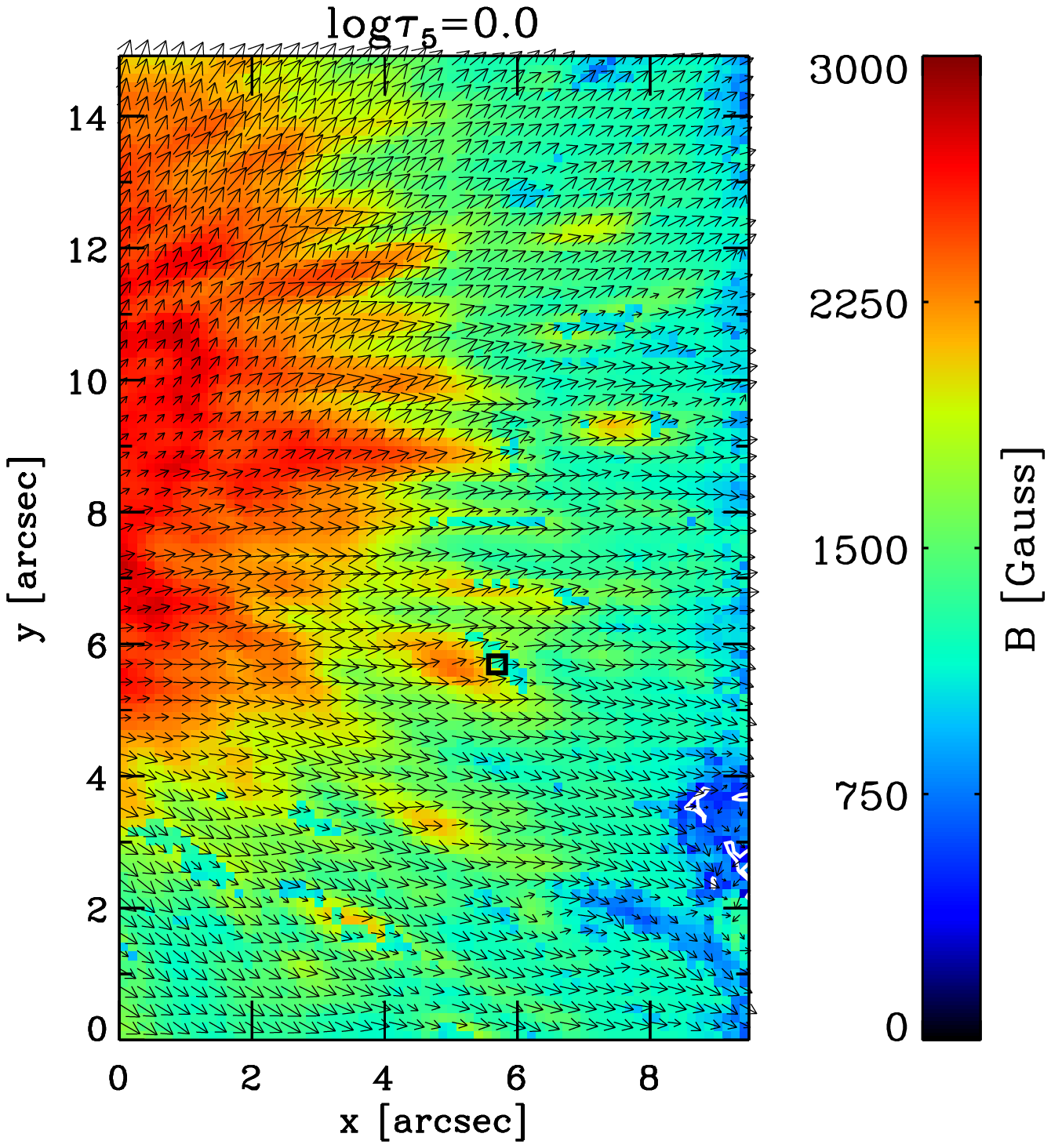} &
\includegraphics[width=9cm]{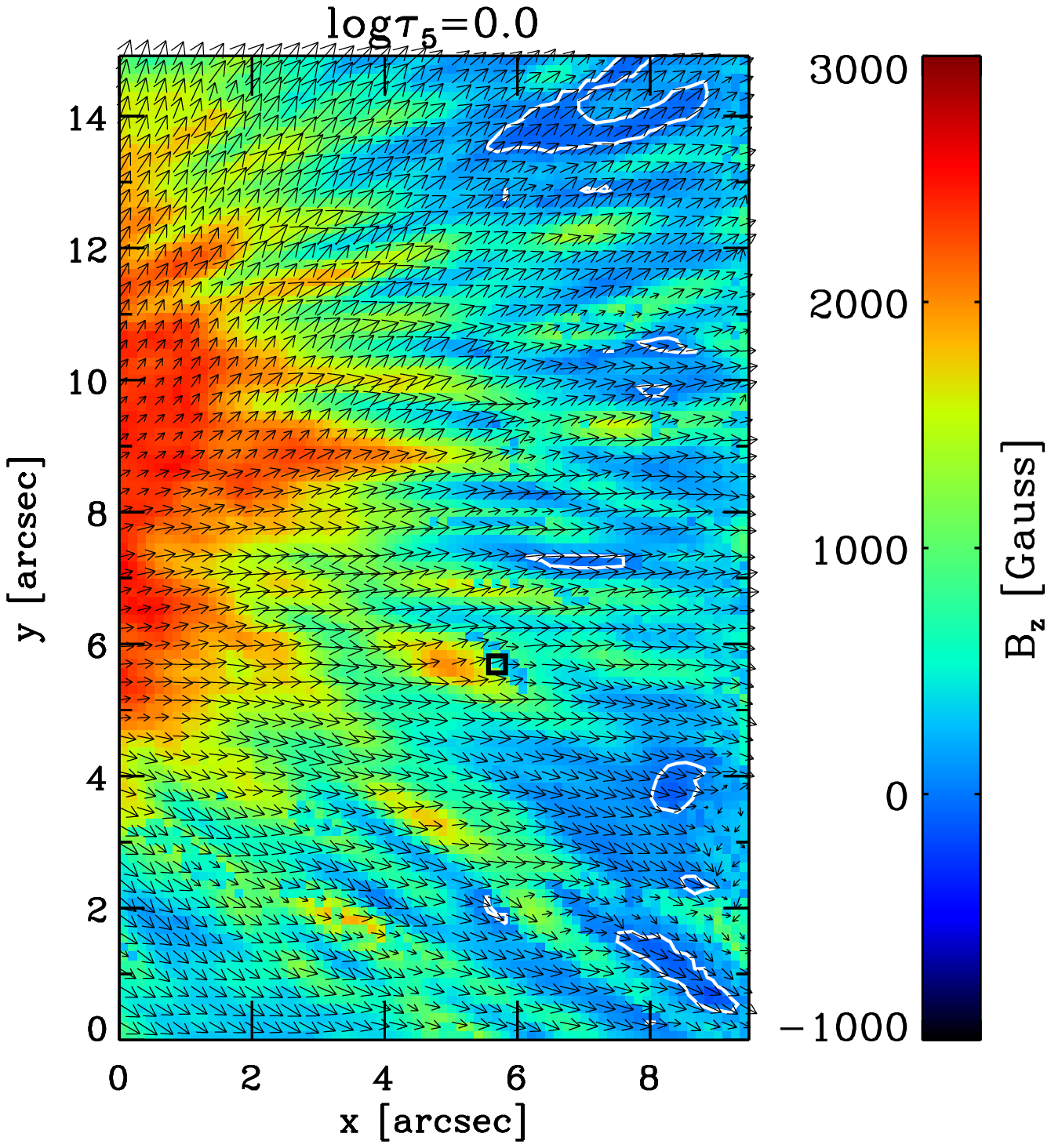} \\
\end{tabular}
\caption{Magnetic field strength $B(x,y,\log\tau_5=0)$ (left) and vertical component
of the magnetic field in the local reference frame $B_z(x,y,\log\tau_5=0)$ (right) in NOAA 12049
observed on May 3rd, 2014 at $\Theta=6.5\deg$. White contours on the left and right panels 
indicate regions where $B<500$ Gauss and $B_z<0$, respectively. Black arrows indicate the projection
of the magnetic field vector $\ve{B}$ on the plane of the solar surface. This region corresponds to 
the red rectangle in Fig.~\ref{figure:ic} (bottom panel).\label{figure:mag_ar12049}. This map was
obtained from the inversion of the deconvolved data assuming $p_w=0.2$ (20 \% wide-angle scattered light)}
\end{center}
\end{figure*}

Figures~\ref{figure:mag_ar12045} and \ref{figure:mag_ar12049} show the maps, at an optical
depth of $\log\tau_5=0$, of the magnetic field strength $B$ (left) and the vertical 
component of the magnetic field $B_z$ in the local reference frame (right) in the regions 
denoted by the red rectangles in Fig.~\ref{figure:ic}, for NOAA 12045 and 12049, respectively.
In these plots the black arrows denote the projection of the magnetic field $\ve{B}$ onto the 
solar surface $(B_r,B_\theta)$. The length of each arrow is proportional to $\sqrt\{B_r^2+B_\theta^2\}$. 
For better visualization we show the vectors only every other pixel both in the vertical and 
horizontal directions. White contours on the $B$ maps (left) enclose those regions where 
$B \le 500$ Gauss, while on the $B_z$ maps (right) they enclose regions where  $B_z \le 0$ (i.e.
magnetic field lines returning to the solar surface).

The most noticeable pattern in these figures is the very well-known spine/intraspine penumbral 
structure \citep{lites1993pen} described in Sect.~\ref{section:intro}: regions of strong and 
vertical magnetic fields (\emph{spines}) interlaced with weaker and more horizontal magnetic 
fields (\emph{intraspines}).

The regions where the magnetic field strength is smaller than 500 Gauss represent only 0.5 \%
and 0.2 \% of the total area in Figures~\ref{figure:mag_ar12045} (left) and \ref{figure:mag_ar12049}
(left), respectively. These areas would be even smaller if we had employed a 300 Gauss limit found by
\cite{spruit2010pen}. At close inspection we notice that these tiny regions where the magnetic
field is below 500 Gauss correspond to pixels in the outer penumbra where granules enter into the sunspot.
A clear example of this can be found in the white contours on the left panel of Fig.~\ref{figure:mag_ar12049} 
($B \le 500$ G) at position $(x \approx 8.4", y \approx 3.7")$. One can see that this region corresponds 
to a bright granule in the outer penumbra inside the red rectangle in the bottom panel of 
Fig.~\ref{figure:ic} at position $(x \approx 53", y \approx 17")$. These results rule 
out the existence, at an optical depth of $\log\tau_5=0$, of dynamically weak magnetic fields 
\citep[cf. ][]{scharmer2008gap,spruit2010pen}, and even more strongly so, the presence of field-free gaps 
in the deep photospheric layers of the penumbra \citep{spruit2006gap,
scharmer2006gap}. This is in agreement with previous results obtained from Hinode/SP data 
\citep{borrero2008penb,puschmann2010pen,basilio2013pen,tiwari2013decon,tiwari2015decon} and SST/CRISP data
\citep{scharmer2008pen,scharmer2012pen,scharmer2013pen}. However, it must be borne in mind that the
information at $\log\tau_5=0$ provided by the spectral lines employed in this work (Sect.~\ref{section:observations}; 
see also Table~\ref{table:atomicdata}) is much more reliable than the information, at the same optical depth, provided 
by the spectral lines (Fe \textsc{i} line pair at 630 nm) employed in the aforementioned investigations (see Section~\ref{subsection:depth}). 
Finally, it is worth mentioning that the results for NOAA 12049 are of particular importance because this sunspot is located 
very close to disk-center, thereby allowing us to probe slightly deeper photospheric layers than NOAA 12045.

The presence of field-free gaps or dynamically weak magnetic fields in the penumbra had been previously ruled out
\citep{mathew2003pen,borrero2004pen,bellot2004pen,borrero2005pen,cabrera2008pen} from observations of the same 
deep-forming Fe \textsc{i} spectral lines around 1565 nm used in this work (see Table~\ref{table:atomicdata}). However,
those older investigations were carried out with data at relatively low spatial resolution (1 arcsec)
and without accounting for wide-angle scattered light.

Regions where $B_z<0$ represent 19.6 \% and 3.0 \%, respectively, of the total area in the
right panels of Figs.~\ref{figure:mag_ar12045} and \ref{figure:mag_ar12049}. The much larger region 
of magnetic flux return in NOAA 12045 is probably due to the presence of a nearby plage with opposite 
polarity magnetic fields outside the sunspot. This also shortens the radial extension of the penumbra 
on the limb side of NOAA 12045 as compared to 12049. These numbers do not fully agree with the values 
given by \citet{basilio2013pen} who found that, at $\log\tau_5=0$, about 28 \% of the penumbra harbors 
magnetic field lines returning into the solar surface. An additional difference difference between our
result and those from \citet{basilio2013pen} and \citet{scharmer2013pen} is that the regions of return 
flux detected here enclose also the central core of the intraspines, not only their lateral boundaries
(see left panels in Figs.~\ref{figure:mag_ar12045} and ~\ref{figure:mag_ar12049}).

The discrepancy in the results might have several sources. One the one hand, as already mentioned 
by \citet[][; see also Sect.~\ref{subsection:depth}]{basilio2013pen}, $B_z$ at $\log\tau_5=0$ is not 
very well constrained by spectropolarimetric observations of the Fe 
\textsc{i} spectral lines at 630 nm (Hinode/SP and SST/CRISP). From this point of view our results at this optical 
depth are to be preferred. On the other hand, as we will show later (see Sect.~\ref{subsection:psfdepen}), 
our results on $B_z$ are heavily dependent on the amount (Eq.~\ref{equation:psf2gauss}; Sect.~\ref{subsection:spatialpsf}) 
of scattered light, $p_w$, employed in the PSF. Since our knowledge of GRIS/GREGOR PSF is very limited 
(Sect.~\ref{section:spectralpsf}) compared to Hinode/SP \citep{suematsu2008hin,danilovic2008psf,vannoort2012decon}, the 
28 \% of return-flux area provided by \citet{basilio2013pen} could be considered as being more reliable. 
Given that both sets of data have their shortcomings we think that the total amount of return flux present in the penumbra and its
spatial distribution is still open to debate. More investigations need to be carried out on this 
subject before a more definite answer can be given.

\subsection{How deep are we probing?}
\label{subsection:depth}

Throughout this paper we have stated several times that the Fe \textsc{i} spectral lines
at 1565 nm employed in this work provide more reliable information about the deep photospheric
layers than the commonly used Fe \textsc{i} spectral lines at 630 nm. In Sect.~\ref{section:observations} 
we ascribed this to the lower H$^{-}$ opacity at 1565 nm than at 630 nm, plus the larger excitation 
potential of the former spectral lines compared to the latter. However this is only a qualitative
explanation. In this section we will provide a more quantitative one. 

\begin{figure}
\begin{center}
\includegraphics[width=9cm]{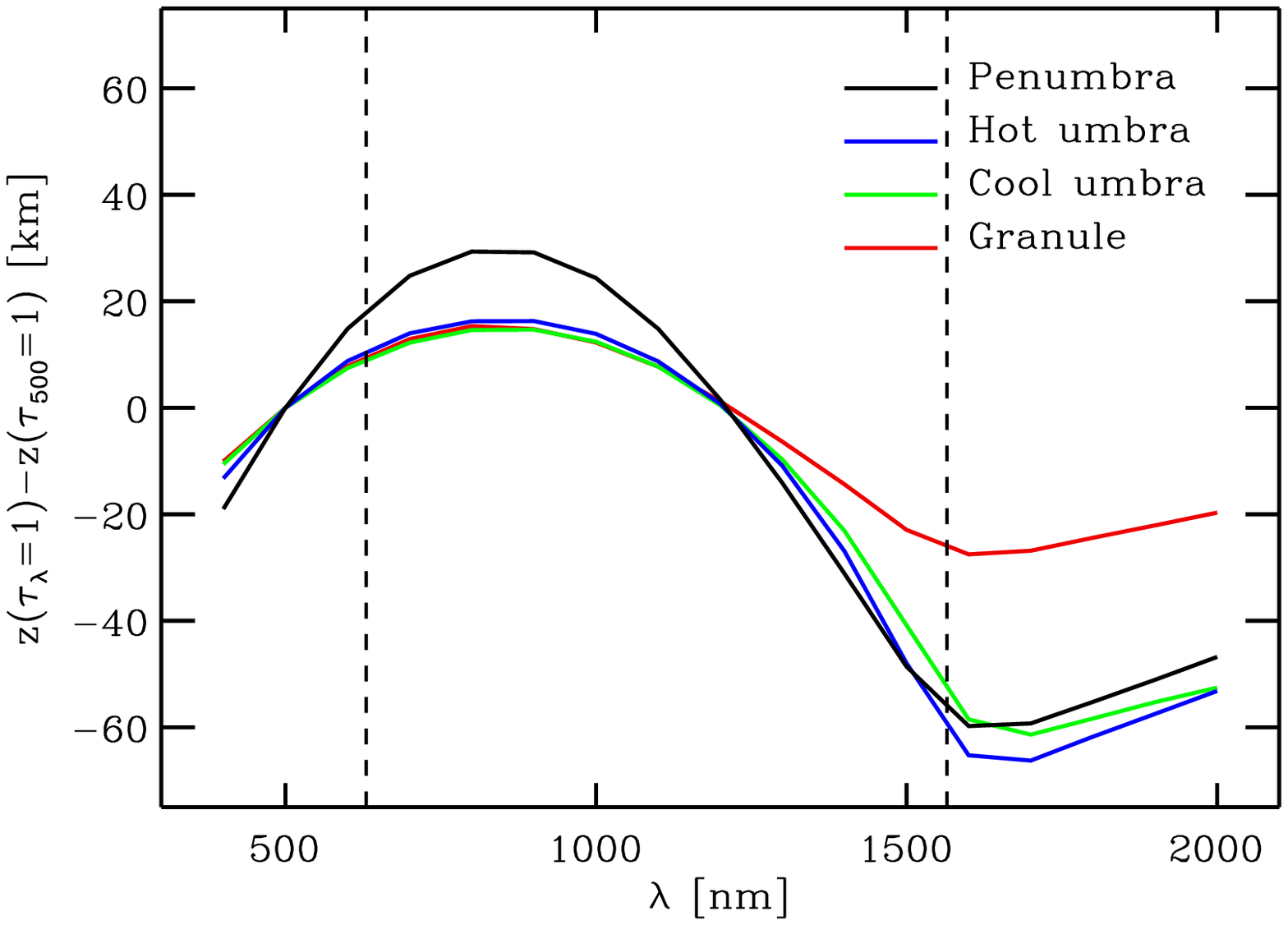}
\end{center}
\caption{Height where the continuum level $z(\tau_{\lambda}=1)$ is formed, with respect to height
at a wavelength of 500 nm, as a function of wavelength. The color curves indicate
the different models employed: red (granular model), blue (hot-small umbral model), green (cool-large
umbral model), black (penumbral model). The two vertical dashed lines are located at $\lambda=630$ nm
and $\lambda=1565$ nm.\label{figure:z_tau_lambda}}
\end{figure}

To that end we have determined the depth of the optical depth unity level as a function of wavelength 
$z(\tau_\lambda=1)$ for different atmospheric models: granular model from \citet{borrero2002gra},
hot and cool umbral models from \citet{collados1994umb}, and finally the spatially-averaged penumbral
model obtained from the inversions of NOAA 12049 described in Sect.~\ref{subsection:inversion}. The results are 
shown in Figure~\ref{figure:z_tau_lambda}. All curves in this figure have been shifted vertically so 
that $z(\tau_\lambda=1)=0$ for a wavelength of 500 nm. As expected, all curves follow the opacity 
due to the bound-bound and bound-free transitions of the H$^{-}$ ion \citep{chandra1946hminus}, but each 
is modulated by the density and temperature of the different models. The height difference between 
the continuum level $\tau_\lambda=1$ or $\log\tau_\lambda=0$ at 630 nm and 1565 nm 
is $-36$ km (granular model; red curve), $-62$ km (cool-large umbra; green curve) and $-71$ km (hot-small umbra; 
blue curve), and $-75$ km (average penumbra; black curve), meaning that the continuum level is formed some 60-70 
km deeper (i.e. half a pressure-scale height) at 1565 nm than at 630 nm in sunspots but only about 30 km deeper 
in the quiet-Sun.

\begin{figure}
\begin{center}
\includegraphics[width=9cm]{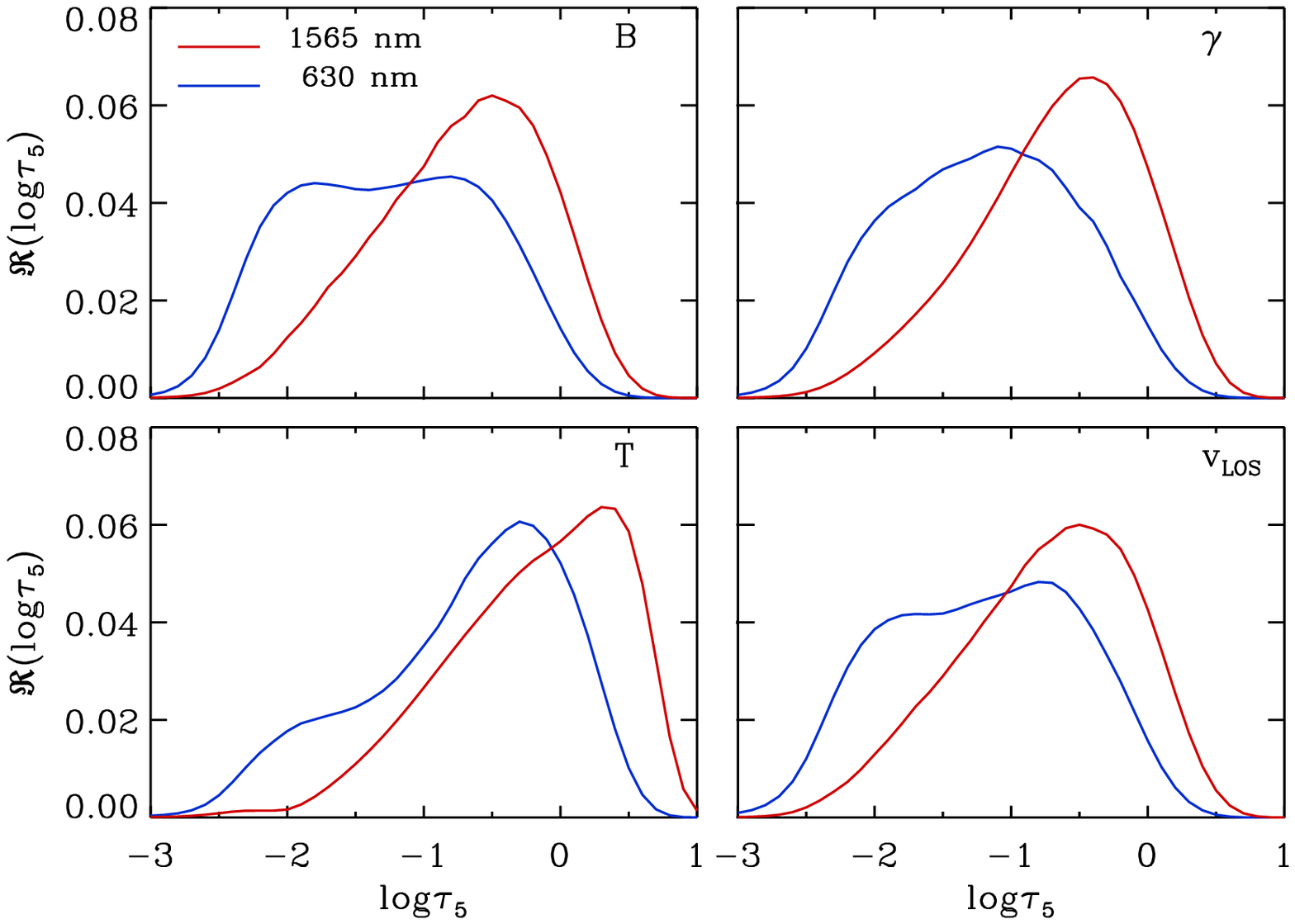}
\end{center}
\caption{Wavelength-integrated response function $\mathcal{R}$
as a function of the logarithm of the optical depth at a reference wavelength of 500 nm, 
$\log\tau_5$, to the magnetic field strength $B$ (top left), magnetic field inclination with
respect to the observer's line-of-sight $\gamma$ (top right), temperature $T$ (lower left), and 
line-of-sight velocity $v_{\rm los}$ (lower right). Red curves correspond to the integration 
over the observed spectral lines in this work (see Table~\ref{table:atomicdata}), whereas
the blue lines correspond to a wavelength integration over the Fe \textsc{i} line pair at 630 nm.
\label{figure:response_penumbra}. The atmospheric model employed in the calculation of the
response functions is the spatially averaged penumbral models obtained from the inversions 
(see Sect.~\ref{subsection:inversion})}
\end{figure}

However, the height of formation of the continuum level tells only part of the story. We must also consider 
the formation of the spectral lines themselves, which depends on their excitation potential, oscillator strength,
and electronic configuration. This can be achieved by means of the so-called response functions $\mathcal{R}(\log\tau_5,\lambda)$ 
\citep{deltoro2003book}. We have calculated the response functions of the spectral lines at 
1565 nm (Table~\ref{table:atomicdata}) and at 630 nm using the average penumbral
model previously mentioned. The contribution from each of the four Stokes parameters
has been taken into account following the method described in \citet{borrero2014milne}. 
Figure~\ref{figure:response_penumbra} presents the wavelength-integrated response functions to the 
magnetic field strength $B$ (top-left), magnetic field inclination with respect to the observer's 
line-of-sight $\gamma$ (top-right), temperature $T$ (bottom-left), and line-of-sight velocity $v_{\rm los}$ 
(bottom-right). These figures demonstrate that the near-infrared (NIR) spectral lines at 1565 observed by GRIS at
the GREGOR telescope and employed in this work are are much more sensitive to the magnetic field $B$ at 
$\log\tau_5=0$ compared to their peak-sensitivity (red curves on the top-left panel) than the visible 
spectral lines at 630 nm observed by Hinode/SP and the CRISP instrument at the SST telescope (blue curves on the 
top-left panel). In particular, the response function to $B$ in the infrared lines at $\log\tau_5=0$ is
about 70 \% of its maximum value, whereas this number decreases to 25 \% for the spectral lines in
the visible. Moreover, the information on the magnetic field conveyed by the 630 nm lines is spread
over a much larger range of optical depths, making it more difficult to isolate the information at 
$\log\tau_5=0$ as compared to the NIR lines, whose contribution comes from a much narrower range.
This suggests that the results obtained at $\log\tau_5=0$ from the inversion of the Fe \textsc{i} lines at 630 nm, 
in particular inversions carried out with two nodes \citep{scharmer2013pen}, tend to be an extrapolation towards 
deeper layers\footnote{This is the case for inversions carried with SIR \citep{basilio1992sir}, or NICOLE
\citep{socas2015nicole}, but not SPINOR \citep{frutiger1999spinor} because the former two codes spread the number
of nodes evenly in $\log\tau_5$ with the first two nodes being located at the top and bottom of the atmosphere.
The latter, however, allows the user to choose their locations.} from the results between 
$\log\tau_5 \in [-2,-1]$. In addition, the peak contribution is much closer to $\log\tau_5=0$ in the NIR 
spectral lines than in the visible ones, with the center-of-gravity 
of $\mathcal{R}$ being located at $\log\tau_5 \approx -1.3$ and at $\log\tau_5 \approx -0.6$, respectively. This implies
that the information about the magnetic field strength $B$ comes from an optical depth of about $10^{-0.6}/10^{-1.3}
\approx 5$ times larger in the case of the NIR spectral lines employed in this work.

\subsection{Dependence on the PSF used to deconvolve}
\label{subsection:psfdepen}

In Sect.~\ref{subsection:spatialpsf} we obtained a simple empirical PSF, $P(x,y)$, for the instrument GRIS
attached to the GREGOR telescope. The PSF was modeled through a narrow- and a wide-angle Gaussian 
profile characterized by $\sigma_n$, $p_n$ and $\sigma_w$, $p_w$ parameters, respectively 
(Eq.~\ref{equation:psf2gauss}). The latter two parameters were rather uncertain, with possible
values such as $\sigma_w \in [10",30"]$ and $p_w\in [0.2,0.3]$. The results presented in 
Sect.~\ref{subsection:fieldfree} (see also Figs.~\ref{figure:mag_ar12045} and \ref{figure:mag_ar12049})
were obtained with $p_w=0.2$, $\sigma_w = 20"$. Motivated by \citet{rolf2013stray} we want to investigate whether our
results depend on the amount of scattered light $p_w$. We have therefore repeated the PCA deconvolution (Sect.~\ref{subsection:pcadecon}) 
and Stokes inversion (as described in Sect.~\ref{subsection:inversion}) using different PSF parameters.
As a first experiment we have simply inverted veil-corrected but un-deconvolved raw data, that is, skipping
Sect.~\ref{subsection:pcadecon} in the analysis. Next, we deconvolved before inversion with the following
combinations: {\bf (a)} $p_w=0$;  {\bf (b)} $p_w=0.2$, $\sigma=20"$; {\bf (c)} $p_w=0.4$, $\sigma=10"$. The other
two parameters were always kept to $p_n=1-p_w$ and $\sigma_n=0.18"$. Table~\ref{table:psfdepen} summarizes our
findings for each experiment in terms of the percentage of the total area covered by regions with $B<500$ Gauss
and $B_z < 0$ at $\log\tau_5=0$. Results for case {\bf (b)} were already presented in Sect.~\ref{subsection:fieldfree}
(see also Figs.~\ref{figure:mag_ar12045} and ~\ref{figure:mag_ar12049}).

\begin{table}
  \begin{center}
    \caption{Percentage of the analyzed penumbral area that harbors weak fields ($B<500$ Gauss)
      or magnetic field lines returning into the solar surface ($B_z<0$) for different {\bf PSF
      parameters}.\label{table:psfdepen}}
  \begin{tabular}{l|cc|cc}
    \hline
    \multirow{2}{*}{Dataset} &
      \multicolumn{2}{c|}{NOAA 12045} &
      \multicolumn{2}{c}{NOAA 12049} \\
    & $B<500$ & $B_z<0$ & $B<500$ & $B_z<0$ \\
    \hline
    \emph{raw} & 0.1\% & 6.1\% & 0.0\% & 0.5\% \\
    \hline
    $p_w=0$ & 0.8\% & 11.9\% & 0.2\% & 0.3\% \\
    \hline
    $p_w=0.2$, $\sigma_w=20"$ & 0.5\% & 19.6\% & 0.2\% & 3.0\% \\
    \hline
    $p_w=0.4$, $\sigma_w=10"$ & 0.1\% & 31.6\% & 0.3\% & 12.7\% \\
    \hline
  \end{tabular}
\end{center}
  \begin{center}
    \caption{Same as Table~\ref{table:psfdepen} {\bf but employing three nodes for $v_{\rm los}$, $B$, $\gamma$, and $\phi$ 
      instead of two nodes}.\label{table:nodedepen}}
  \begin{tabular}{l|cc|cc}
    \hline
    \multirow{2}{*}{Dataset} &
      \multicolumn{2}{c|}{NOAA 12045} &
      \multicolumn{2}{c}{NOAA 12049} \\
    & $B<500$ & $B_z<0$ & $B<500$ & $B_z<0$ \\
    \hline
    \emph{raw} & 0.0\% & 6.0\% & 0.0\% & 0.5\% \\
    \hline
    $p_w=0$ & 0.3\% & 12.6\% & 0.5\% & 0.0\% \\
    \hline
    $p_w=0.2$, $\sigma_w=20"$ & 0.3\% & 27.8\% & 0.1\% & 2.3\% \\
    \hline
    $p_w=0.4$, $\sigma_w=10"$ & 0.1\% & 48.4\% & 0.2\% & 10.2\% \\
    \hline
  \end{tabular}
\end{center}
\end{table}

Clearly the penumbral area featuring regions where the magnetic field returns to
the solar surface ($B_z<0$) strongly depends on the amount of scattered light $p_w$ employed to model
the PSF. As $p_w$ increases the areas harboring return flux becomes larger. The area
covered by weak magnetic fields ($B<500$) is however independent of $p_w$.

\subsection{Dependence on the number of nodes employed in the inversion}
\label{subsection:nodedepen}

So far, the results presented in this paper were obtained employing, during the inversion process,
three nodes in the temperature $T$, and two nodes in the line-of-sight velocity $v_{\rm los}$, magnetic field
strength $B$, inclination of the magnetic field with respect to the observer's line-of-sight $\gamma$,
and angle of the magnetic field in the plane perpendicular to the observer's line-of-sight $\psi$ 
(see Section~\ref{subsection:inversion}). Using two nodes in the aforementioned physical parameters 
assumes that each of them varies linearly with the logarithm of the optical depth $a+b*\log\tau_5$, 
with the slope $b$ and zero crossing $a$ being different for $B$, $\gamma$, etc. Depending on the sign
of the slope this implies that a given physical parameter can either increase or decrease with height in
the atmosphere. A slightly more realistic, albeit complex, situation would be to allow for three nodes instead
of two (i.e. quadratic dependence of the physical parameters with $\log\tau_5$) and hence allowing them
to first increase with height and then decrease, or vice-versa. To determine how our results depend on the choice
of nodes we have repeated all inversions in Sect.~\ref{subsection:psfdepen} but employing three nodes in
$v_{\rm los}$, $B$, $\gamma$, and $\phi$. Table~\ref{table:nodedepen} shows the percentage of the penumbral
area harboring weak fields ($B<500$ Gauss) and field lines returning to the solar surface ($B_z<0$) employing
three nodes. Comparing Table~\ref{table:nodedepen} with Table~\ref{table:psfdepen} we conclude that the size
of flux-return areas significantly depend on whether each physical parameter is modeled with two or three nodes. 
The area of weak fields, however, are the same in both cases.

\section{Conclusions}%
\label{section:conclu}

We have studied the magnetic field topology in the penumbra of two sunspots 
at the deepest layers of the solar photosphere. This was done through the inversion
of the radiative transfer equation applied to spectropolarimetric data 
(i.e. full Stokes vector $\ve{I}$) of three Fe \textsc{i} spectral lines at 1565 nm
in order to retrieve the magnetic field $\ve{B}$. 

The estimated spatial resolution of the data employed in work is 0.4-0.45\arcsec\ and the 
noise level is $10^{-3}$. Moreover, the observed spectral lines
are better suited to study the magnetic field in the deep photosphere than the widely
used Fe \textsc{i} spectral lines at 630 nm because, besides the Zeeman
splitting being about three times larger than in the lines at 630 nm, the lines
at 1565 nm convey information from deeper photospheric layers.

In order to account for the degradation of the data due to straylight (i.e.
wide-angle scattered light) within the instrument, we have applied, prior
to the inversion, a PCA deconvolution method using an empirical point
spread function. Our results show no evidence for the presence of weak-field 
regions ($B<500$), let alone of dynamically weak fields \citep{spruit2010pen} or 
field-free regions \citep{scharmer2006gap,spruit2006gap} in the deepest regions 
of the photosphere ($\log\tau_5=0$). This agrees with previous observational results, in particular 
with \citet{borrero2008penb,borrero2010pen,tiwari2013decon} and with three-dimensional
MHD simulations of sunspot fine-structure \citep{rempel2009mhd,nordlund2010mhd,rempel2011mhd,rempel2012mhd}.
These results are independent from the amount of straylight used in the PSF and independent
of the inversion set-up (i.e. number of nodes). None of the aforementioned works can rule
out the existence of field-free plasma deep beneath the sunspot. Indeed it is perfectly plausible that
at some point underneath the sunspot \citep[i.e. under the magneto-pause;][]{jahn1994sun} normal
field-free convection resumes. The question is wether this happens sufficiently close to $\log\tau_5=0$
so as to explain the penumbral brightness. This is precisely what our work rules out.

On the other hand, the amount of flux returning 
back into the solar surface ($B_z<0$) within the penumbra is very dependent on the amount of 
straylight considered and, consequently, we shall refrain from drawing conclusions at this point.

In summary, we have addressed all major concerns raised by \citet{spruit2010pen}, \cite{scharmer2012pen},
and \citet{scharmer2013pen}: we have used high-spatial resolution observations (indeed the highest ever 
at 1565 nm) of spectropolarimetric data \citep{scharmer2012pen} that conveys very reliable 
information about the magnetic field in the deep Photosphere \citep{spruit2010pen}. We have also deconvoled the data with 
several empirically determined PSFs, thereby allowing us to remove the need for the so-called non-magnetic 
filling factor \citep{scharmer2013pen}. In all cases, no traces of regions with where $B \le 500$ Gauss 
have been found at $\log\tau_5=0$. 

\begin{acknowledgements}
The 1.5-meter GREGOR solar telescope was built by a German consortium under the
leadership of the Kiepenheuer-Institut für Sonnenphysik in Freiburg with the
Leibniz-Institut für Astrophysik Potsdam, the Institut für Astrophysik
Göttingen, and the Max-Planck-Institut für Sonnensystemforschung in Göttingen as
partners, and with contributions by the Instituto de Astrofísica de Canarias and
the Astronomical Institute of the Academy of Sciences of the Czech Republic.
We are very grateful to the engineering, operating and technical staff at the GREGOR Telescope: 
Andreas Fischer, Olivier Grassin, Roberto Simoes, Clements Halbgewachs, Thomas Kentischer, 
Thomas Sonner, Peter Caligari, Michael Wiesssch\"adel, Frank Heidecke, Stefan Semeraro, and 
Oliver Wiloth. This research has made use of NASA's Astrophysics Data System. 
\end{acknowledgements}

\bibliographystyle{aa}
\bibliography{borrero_aa28313}

\end{document}